\documentclass[]{aa}
\usepackage{graphicx}
\usepackage{hyperref}
\usepackage{natbib}
\usepackage{color}
\usepackage[utf8]{inputenc}
\usepackage[varg]{txfonts}
\usepackage{mathabx}
\usepackage{multirow}
\usepackage{placeins}
\usepackage{amsmath}
\usepackage{chemformula}

\usepackage{xspace}
\newcommand{\Rearth}{$R_{\oplus}$\xspace}

\makeatletter
\renewcommand*{\@fnsymbol}[1]{\ensuremath{\ifcase#1\or \bigstar\or \bigstar\bigstar\or \ddagger\or
   \mathsection\or \mathparagraph\or \|\or **\or \dagger\dagger
   \or \ddagger\ddagger \else\@ctrerr\fi}}
\makeatother

\usepackage[free-standing-units=true]{siunitx}
\newcommand{\micron}{\,\si{\micro\meter}}

\usepackage{hyperref}
\hypersetup{
    colorlinks = true,
    linkcolor = {blue}, 
    citecolor = {blue},
    urlcolor = {blue}
}

\begin{document}
\nolinenumbers
\title{Competing chemical signatures in the atmosphere of TOI-270 d:\\Inference of sulfur and carbon chemistry}
    
\titlerunning{Competing chemical signatures in the atmosphere of TOI-270 d}
\author{
L. Felix\inst{1,2}, $^{\href{https://orcid.org/0009-0004-1292-3969}{\includegraphics[scale=0.008]{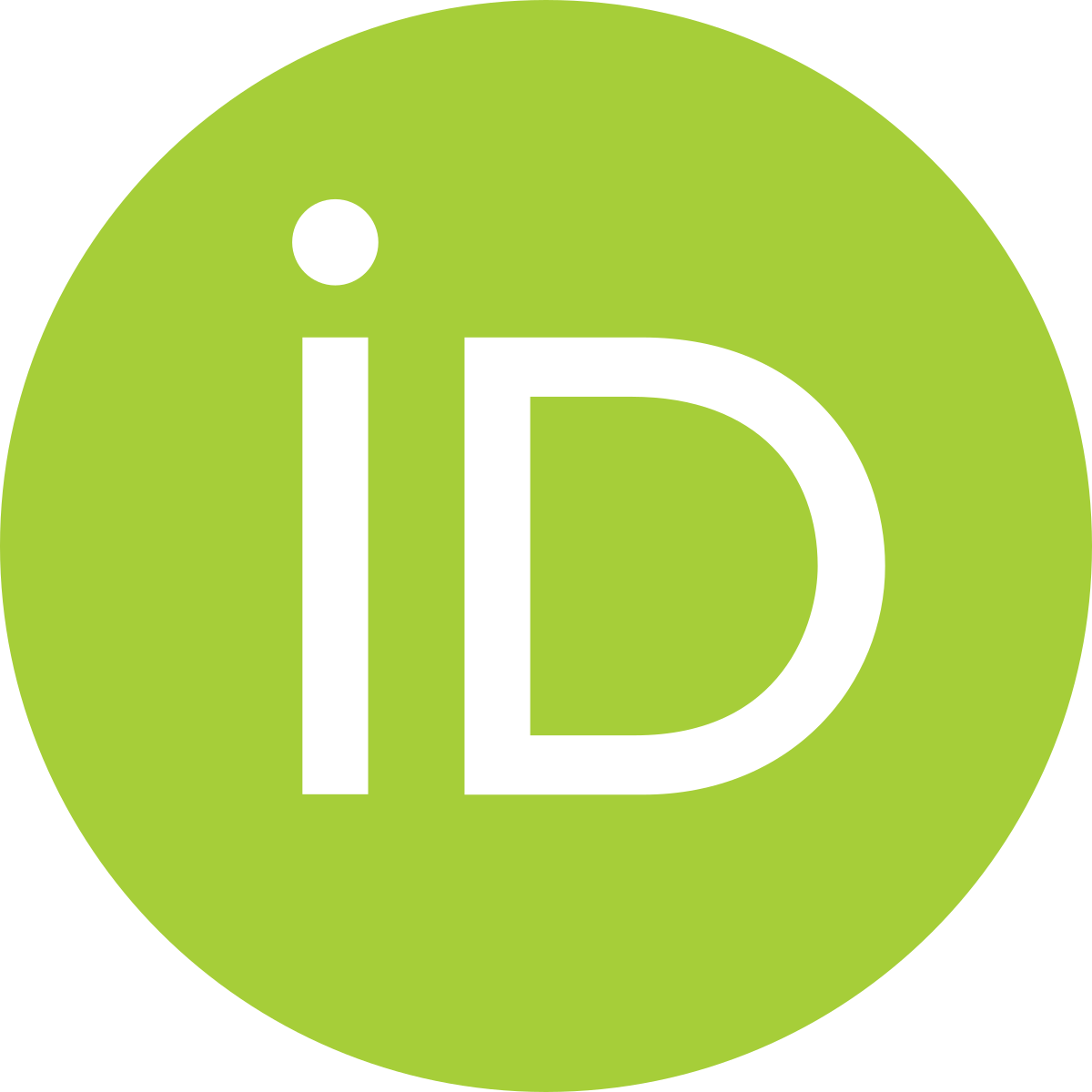}}}$,
D. Kitzmann\inst{3,2} $^{\href{https://orcid.org/0000-0003-4269-3311}{\includegraphics[scale=0.008]{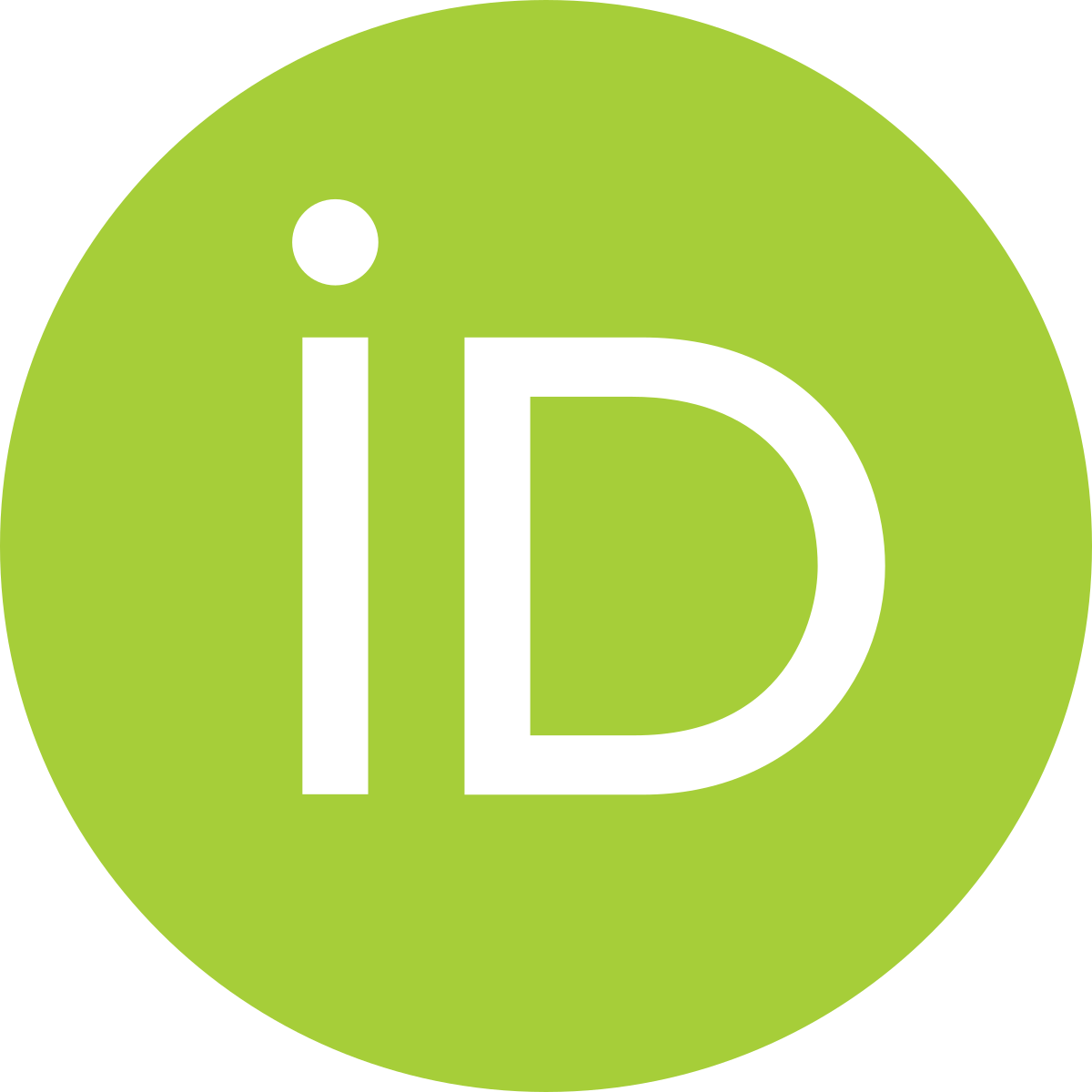}}}$,
B.-O. Demory\inst{2,3,4} $^{\href{https://orcid.org/0000-0002-9355-5165}{\includegraphics[scale=0.008]{figures/orcid.png}}}$,
C. Mordasini\inst{3,2} $^{\href{https://orcid.org/0000-0002-1013-2811}{\includegraphics[scale=0.008]{figures/orcid.png}}}$
}

\institute{
\label{inst:1} Institute for Particle Physics \& Astrophysics, ETH Zurich, Wolfgang-Pauli-Str. 27, 8093 Zurich, Switzerland \and
\label{inst:2} Center for Space and Habitability, University of Bern, Gesellschaftsstrasse 6, 3012 Bern, Switzerland \and
\label{inst:3} Division of Space Research and Planetary Sciences, Physics Institute, University of Bern, CH-3012 Bern, Switzerland \and
\label{inst:4} ARTORG Center for Biomedical Engineering Research, University of Bern, Murtenstrasse 50, CH-3008, Bern, Switzerland
}
              
\authorrunning{L. Felix et al.}
\date{Received: 17 April 2025 / Accepted:  29 July 2025}

\abstract   
{Recent JWST measurements allow access to the near-infrared spectrum of the sub-Neptune TOI-270 d, for which two different interpretations, a high-metallicity miscible envelope and a lower metallicity hycean world, are currently in conflict.}
{Here, we reanalyze published data and reproduce previously retrieved molecular abundances based on an independent data reduction and a different retrieval framework. The aim of this study is to refine the understanding of TOI-270 d and highlight the impact of various choices during JWST data analysis. Particularly, we test the impact of data resolution on atmospheric retrieval calculations.}
{We reduced one JWST NIRSpec G395H and one NIRISS SOSS GR700XD transit dataset using the Eureka! pipeline and a custom Markov Chain Monte Carlo-based light curve fitting algorithm at the instruments' native resolutions. The atmospheric composition was estimated with the updated \textsc{BeAR} retrieval code across a grid of retrieval setups and spectral resolutions.}
{Our transit spectrum is consistent with previous studies except at the red end of the NIRISS data. Our retrievals support a \ch{H2}/He-dominated atmosphere with high mean molecular weight for TOI-270 d. We provide refined abundance constraints and find statistically favored model extensions indicating either sulfur-rich chemistry with species such as \ch{CS2}, \ch{CS}, and \ch{H2CS} or the possible presence of \ch{CH3Cl} or \ch{CH3F}. However, Bayesian inference cannot distinguish between these scenarios due to similar opacities below $4\micron$. To obtain physically plausible atmospheric solutions at native resolution, accounting for the instrument's line spread function is essential.}
{Our analysis reinforces TOI-270 d as a highly interesting warm sub-Neptune for atmospheric studies, with a complex chemistry in a cloud-free upper atmosphere. However, its exact nature remains uncertain and warrants further detailed photochemical modeling and observations.}


\maketitle

\section{Introduction} \label{section:introduction}
\nolinenumbers
The radius distribution of small exoplanets has been found to be bimodal in nature \citep{Fulton_2017}, with a minimum in planet numbers that have radii around $1.7-2.0\,R_\oplus$.
The smaller end of this distribution is typically thought to be super-Earths, rocky planets with little to no atmospheres \citep{owen_atmospheric_2019}.
The nature of the other end of the population ($R_\mathrm{p} < 4 R_\oplus$), usually called sub-Neptunes, still remains largely unclear, even though they are the most common type of planet \citep{annurev_stats_2021, gupta_sculpting_2019}.
The lack of a solar system analog combined with the mass degeneracy of various possible chemical compositions \citep{Rogers_interiordegeneracy_2010, mousis_irradiated_2020, Dorn_hiddenwater_2021} has led to a range of possible planet structures. 
These vary from small gas planets, so-called mini-Neptunes \citep{wogan_jwst_2024}, to rocky bodies with low-metallicity atmospheres, sometimes referred to as hyterran worlds \citep{emsenhuber_popsynth_rev_2023}. In between these two extremes are other potential planetary structures with varying ratios of core, water, and envelope mass fractions.
Examples for such planets have been described as hycean worlds by \citet{madhusudhan_habitability_2021}, as magma ocean worlds \citep{massol_magmaocean_atmospheres_2016}, or steam worlds \citep{Zeng_waterworlds_2019, burn_radius_2024}.

With the James Webb Space Telescope (JWST) \citep{gardner_jwst}, we are now entering an era where the detailed characterization of these worlds has become accessible, allowing us to search for and probe the predicted chemical diversity of sub-Neptunes \citep{Fortney_subnep_characterization_2013, Hu_shroudedoceans_2021, guzman-mesa-chemical-2022}. A couple of sub-Neptunes have already been observed in both transit and occultation using the JWST with varying degrees of success.
One prominent JWST target is K2-18 b ($R_\mathrm{p}=2.6\,R_\oplus$, \citet{benneke_k21-8b_scarlet_2019}). The obtained spectra, with detections of methane (\ch{CH4}) and carbon dioxide (\ch{CO2}), as well as as a claimed potential presence of dimethyl sulfide (DMS) \citep{madhusudhan_carbon_2023}, indicated that this cold sub-Neptune (equilibrium temperature $T_\mathrm{eq}$ = 255 K) could be a potential hycean planet. The atmospheric composition of K2-18 b is still heavily debated, though (e.g. \citet{wogan_jwst_2024}, \citet{Cooke_photochemicalmodelling_2024}). For example, a recent reanalysis of the JWST data has only been able to confirm the presence of \ch{CH4} at a high enough significance for detection \citep{schmidt_k2-18b_comprehensive_2025}.
Another sub-Neptune, GJ 3470 b ($R_\mathrm{p}=4.2\,R_\oplus,\,T_\mathrm{eq}=600\,\text{K}$, \citet{bonfils_gj3470b_detection_2012}), has reported detections of water (\ch{H2O}), \ch{CO2}, \ch{CH4}, and sulfur dioxide (\ch{SO2}), suggesting an atmosphere with elemental abundances roughly one hundred times those of the sun \citep{beatty_gj3470b_2024}. GJ 3470 b is currently also the coldest exoplanet with a constraint on the presence of sulfur dioxide, which would require a significant contribution by disequilibrium chemistry.
Analysis of JWST data of GJ 9827 d ($R_\mathrm{p} = 2.0\,R_\oplus,\,T_\mathrm{eq} = 620\,\mathrm{K}$, \citet{prieto_gj9827d_det_2018}) by \citet{Piaulet_gj9827_2024} put a constraint of  roughly $32\%$ on the atmospheric water mass fraction, which led to the conclusion that this planet is likely a so-called steam world. \citet{davenport_toi-421b_2025} analyzed observations of TOI-421 b ($R_\mathrm{p}=2.6\,R_\oplus,\,T_\mathrm{eq}=920\,\text{K}$, \citet{Krenn_toi421b_char_2024}). Their atmospheric retrieval analysis resulted in robust detections of \ch{H2O} with signs of \ch{CO} and \ch{SO2} and a mean molecular weight of less than 2.5 amu.

TOI-270 d, also known as L-231-32 d, was initially discovered in 2019 by \citet{gunther_toi270_2019} through transit observations with the Transiting Exoplanet Survey Satellite (TESS) around a relatively calm M3.0V star. Its radius of $2\,R_\oplus$ puts this planet into the sub-Neptune regime. Follow-up radial velocity measurements by \citet{vaneylen_toi270_2021} provided the first mass constraints of $4.78\pm0.43\,M_\oplus$, leading to an initial density measurement of $2.72\pm0.33\,\text{g/cm}^3$ for TOI-270 d. Assuming a Bond albedo of 0.3, the estimated equilibrium temperature of TOI-270 d would be $T_\mathrm{eq} = 340$ K. 
\citet{mikalevans_toi270_2023} conducted a first transit spectroscopy study using data from the Wide Field Camera 3 (WFC3) of the Hubble Space Telescope (HST). 
They concluded that an atmosphere rich in molecular hydrogen (\ch{H2}) together with water as a main absorber is the best explanation for their transmission spectrum.

TOI-270 d is considered an ideal sub-Neptune target for atmospheric characterization with JWST. Therefore, the planet has been the target of multiple General Observer (GO) programs (GO programs 2759, 3557, and 4098). 

In this study, we focus on the first transit observations of TOI-270 d, part of the GO program 4098 (PI Benneke \& Evans-Soma). Through this program, transits were observed with the Near Infrared Imager and Slitless Spectrograph (NIRISS) and the G395H grism of the Near Infrared Spectrograph (NIRSpec). Combined, both instruments offer a wavelength range from 0.6 $\mu$m to 5.2 $\mu$m, which allows for the detection of many important molecules.
Two teams have previously analyzed JWST datasets of this planet: \citet{holmberg_possible_2024} used a combination of the JWST NIRSpec data and the WFC3 spectrum by  \citet{mikalevans_toi270_2023}, while \citet{benneke_jwst_2024} focused on the NIRISS and NIRSpec observation from the GO program 4098.
\citet{benneke_jwst_2024} reported an atmosphere dominated by molecular hydrogen (\ch{H2}) and helium (\ch{He}) that is well mixed with heavier molecules and concluded that TOI-270 d is a metal-rich miscible-envelope sub-Neptune. They constrained a mean molecular weight of about 5.5 amu and claimed detections of \ch{CH4} and \ch{CO2}, and minor evidence for the presence of \ch{H2O} and carbon disulfide (\ch{CS2}). In contrast, \citet{holmberg_possible_2024} described TOI-270 d as a potential hycean planet based on their separate data reduction and atmospheric retrieval calculations.
They too provided detections for \ch{CH4} and \ch{CO2}, with minor constraints on \ch{H2O} and \ch{CS2}.

In this work we focus on the sub-Neptune TOI-270 d and use the existing JWST data sets of the GO program 4098. We performed a new, independent data reduction of the available NIRISS and NIRSpec observations, which we describe in Sect. \ref{section:reduction}. With the obtained transmission spectra, we then performed atmospheric retrieval calculations to constrain the composition and temperature of the planet's atmosphere. The results are discussed in Sect. \ref{section:retrieval_results}. We gave particular attention to investigating the impact of spectral data binning on the retrieval results and tested possible extensions of the forward model. Conclusions and a summary are provided in Sect. \ref{section:discussion}.

\section{JWST observations and data reduction}
\label{section:reduction}

In this section, we describe the data reduction procedure to transform the original JWST raw data to transmission spectra. We describe the reductions of the two different sets from NIRSpec and NIRISS separately. We note that in addition to the NIRSpec G395H transit observation used here, an additional data set with the same instrument was obtained through the GO program 4098. Since the latter data has not been officially published yet, we focus on the single observation used in \citet{benneke_jwst_2024} and \citet{holmberg_possible_2024}.

\subsection{NIRSpec G395H observation}
The NIRspec dataset used in this work was obtained on 4 October 2023 through the GO 3557 and 4098 programs. 
It consists of 1763 integrations with 11 groups each, totaling an exposure time of 4.9 hours within the 5.3 hour observation window. 
The observations were taken in the bright-object-timeseries (BOTS) mode using the F290LP filter and G395H grating and covering $2.74-5.17\micron$.
The chosen readout pattern is NRSRAPID on the SUB2048 subarray of the detector.
With the G395H grating, the spectrum is projected onto two separate detectors, NRS1 and NRS2, leading to a gap of $\sim0.1\micron$ between $3.7-3.8\micron$. 
Notably, this observation also includes a transit of the rocky planet TOI-270 b, which overlaps at the beginning of the transit of TOI-270 d.

\begin{figure*}[t]
    \centering
    \resizebox{0.48\hsize}{!}{\includegraphics{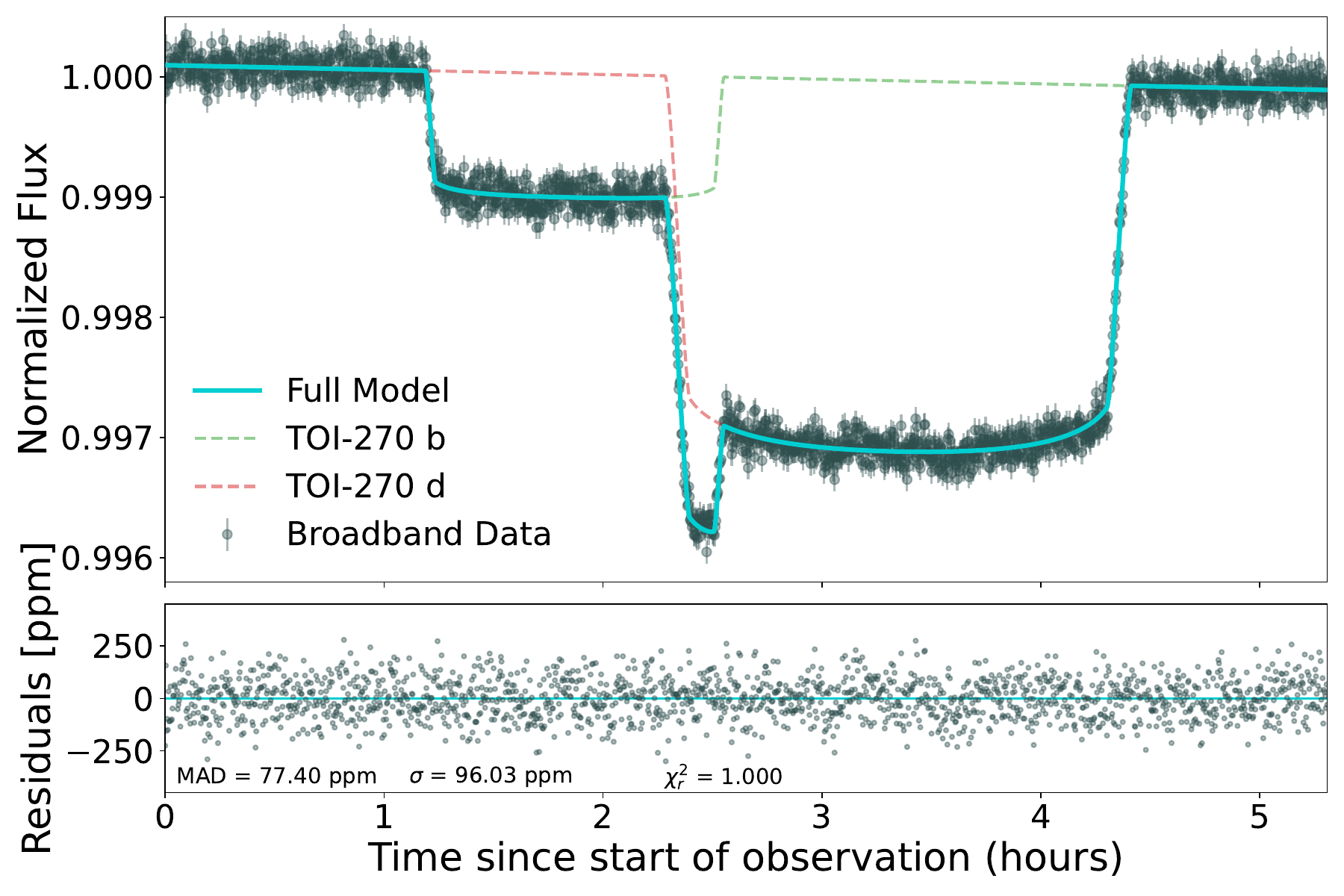}}
    \resizebox{0.48\hsize}{!}{\includegraphics{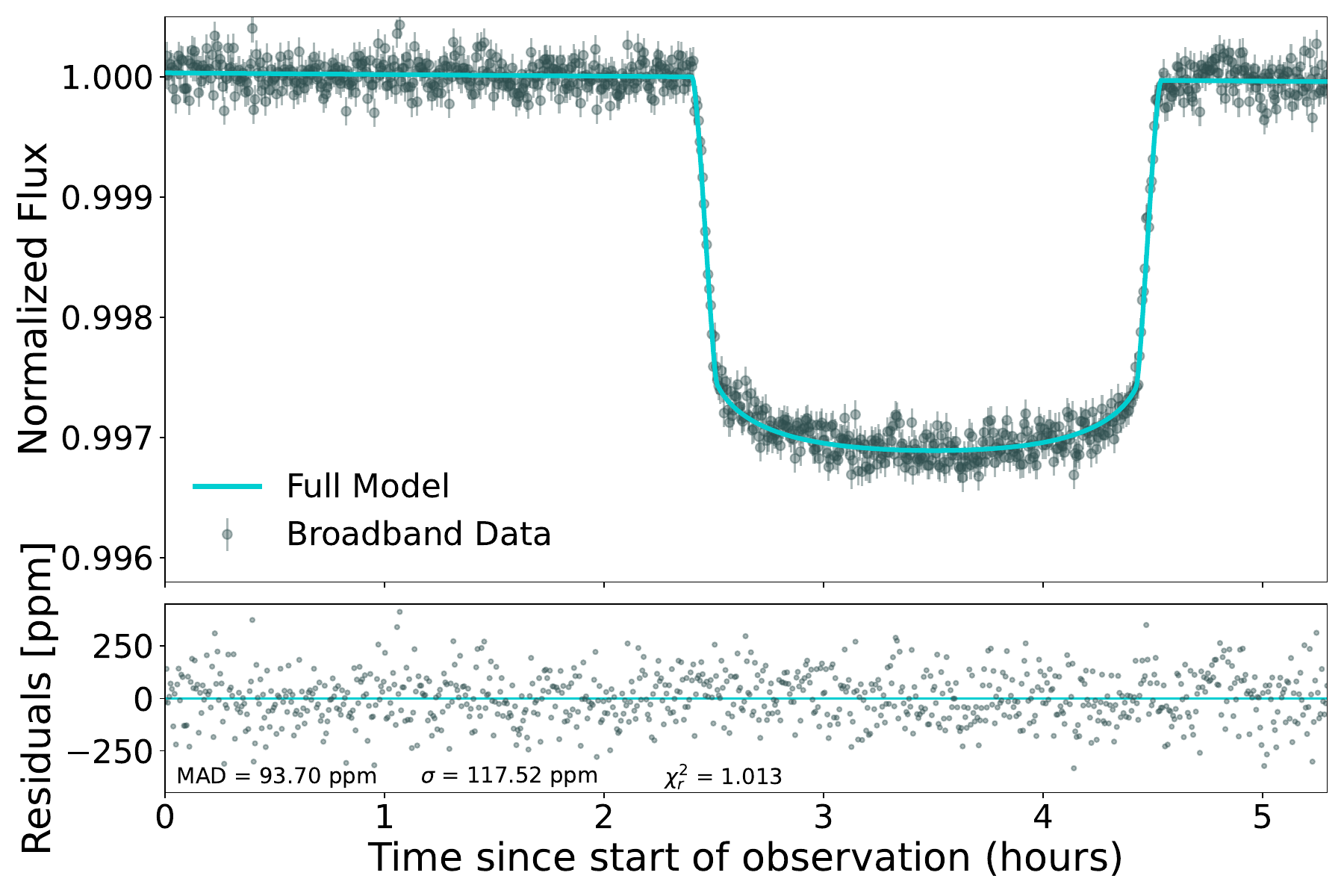}}
    \caption{Broadband light curve fits for both JWST observations. \textbf{Left:} Combined NIRSpec NRS1 and NRS2 broadband light curve fit of the overlapping transits of TOI-270 b and TOI-270 d. JWST ETC calculations predict noise levels of $\approx85$ ppm, which is roughly 12\% below the standard deviation of our residuals. \textbf{Right:} NIRISS first order broadband light curve fit of the transit of TOI-270 d. From S/N calculations with the ETC one would expect a noise level of $\approx69$ ppm, much lower than we find to be the case for the broadband light curve. This could partially be caused by stellar activity, see \ref{app:niriss_order2}.}
    \label{fig:broadband_lcs}%
\end{figure*}

\subsubsection{Eureka! reduction}
We reduced the NIRSpec observation with the commonly used Eureka! pipeline \citep{bell_eureka_2022}. However, we only used the first three stages of Eureka! before we continued with our own custom light curve fitting procedure. Stages 1 and 2 of Eureka! act as a wrapper for the standard JWST data reduction pipeline \citep{bushouse_2023}, with the third stage covering the spectral extraction by using an optimal extraction algorithm \citep{horne_optspec_1986}. In the following, we summarize some of the choices we made during the data reduction. 
Our input .ecf-files for Eureka! are available on zenodo\footnote{\url{https://doi.org/10.5281/zenodo.15228454}}, but for convenience, a table summarizing the most important settings can be found in Appendix \ref{section:eureka}.

We started with the uncalibrated JWST data products and made some important changes compared to the default Eureka! settings. For the jump-detection step, we increased the detection threshold to 8-$\sigma$ to avoid correcting false positives \citep{Alderson_earlynirspec_2023}. It has been noted by, for example, \citet{moran_jwstgj486b_2023} or \citet{alderson_jwst_2024}, that the bias-correction step can improve photometric scatter at the cost of introducing detector offsets. 
In this work, we therefore chose to optimize for photometric scatter by using a smooth bias correction based on the first group, with a window length of roughly 30 minutes. Since we aimed to combine data from different detectors, which already introduces potential offsets, using the bias correction here therefore did not introduce additional artifacts. 

From the first stage of Eureka!, we chose to skip two particular steps: the IPC-step (which is skipped by default) and the persistence correction, which is typically not used for near-infrared time-series observations. One important step in the first stage of Eureka! is the additional group-level $1/f$-correction \citep{Rustamkulov_2022}. This correction is simply a column-wise subtraction of the mean background-pixel count, calculated from all pixels in the column that are outside the masked trace (9 pixels wide) and not rejected by a sigma clipping. In our reduction, we found that a fairly strict clipping threshold of 2-$\sigma$ was needed to avoid overcorrecting backgrounds by values of the order of 500 $e^-$. Using this rejection threshold yielded typical background values of around 25-50 $e^-$. 

In the second stage of Eureka!, we skipped all steps except for \texttt{srctype}, \texttt{wavecorr}, \texttt{extract1d}, and \texttt{extract2d}. All other corrections are not needed for the relative flux measurements of exoplanet transit observations and could potentially bias our resulting spectrum.

During the third stage, Eureka! shifts the position of the pixels vertically in order to align the spectral trace, using a Gaussian fit of the measured flux. Following that step, we then applied another background-subtraction, considering only the data that is at least 7 pixels away from the center of the trace. For this second background subtraction a clipping threshold of $3\sigma$ was found to be sufficient to avoid overcorrections. This yields corrections of up to 25 $e^-$ for NRS1 and up to 30 $e^-$ for NRS2. Finally, the optimal spectrum extraction \citep{horne_optspec_1986} was executed with an aperture half-width of 4 pixels for NRS1 and 3 pixels for NRS2. These half-widths have been chosen such that they minimize themedian absolute deviation (MAD) of the resulting spectra.

\begin{table*}[t!]
\renewcommand*{\arraystretch}{1.3}
\caption{Priors and posteriors for the broadband MCMC light curve fit of the NIRspec BOTS G395H and NIRISS SOSS data, as shown in Fig. \ref{fig:broadband_lcs}. 
}
\label{table:broadband_results}
\resizebox{\hsize}{!}{
\begin{tabular}{lcccc}
\hline \hline
Parameters                                       & \multicolumn{2}{c}{NIRSpec}                                                 & \multicolumn{2}{c}{NIRISS}                                                  \\
                                 & Prior                             & Broadband Fit                           & Prior                             & Broadband Fit                           \\ \hline
\textbf{TOI-270 d}                                  &                                   &                                         &                                   &                                         \\
$t_{0}$ (BJD)      & $\mathcal{U}(60221.26, 60221.41)$ & $60221.3017 \pm 2\cdot10^{-5}$ & $\mathcal{U}(60346.43, 60346.53)$ & $ 60346.478015 \pm 3.5\cdot10^{-5}$ \\
$R_\mathrm{p}/R_*$                  & $\mathcal{U}(0.04, 0.06)$         & $0.053987^{+0.000067}_{-0.000074}$      & same as NIRSpec                   & $0.05362 \pm 0.00013$           \\
$a/R_*$                  & $\mathcal{U}(40, 44)$             & $42.73^{+0.37}_{-0.25}$                 & same as NIRSpec                   & $42.33^{+0.67}_{-0.52}$                 \\
$i$ ($\deg$) & $\mathcal{G}(89.73, 0.16)$        & $89.833^{+0.078}_{-0.084}$              & same as NIRSpec                   & $89.75 \pm 0.12$                 \\
\hline
\textbf{TOI-270 b}                                  &                                   &                                         &                                   &                                         \\
$t_{0}$ (BJD)      & $\mathcal{U}(60221.16, 60221.31)$ & $60221.2401 \pm 3.5\cdot10^{-5}$  & -                                 & -                                       \\
$R_\mathrm{p}/R_*$                  & $\mathcal{U}(0.02, 0.04)$         & $0.0315356 \pm 9.9\cdot10^{-5} $     & -                                 & -                                       \\
$a/R_*$                  & $\mathcal{U}(16, 20)$             & $18.79^{+0.58}_{-0.45}$                 & -                                 & -                                       \\
$i$ ($\deg$) & $\mathcal{G}(89.39, 0.37)$        & $89.19^{+0.31}_{-0.34}$                 & -                                 & -                                       \\
\hline
\textbf{Other parameters}                   &                                   &                                         &                                   &                                         \\
$u_1$                         & $\mathcal{G}(0.037, 0.050)$        & $0.104 \pm 0.019$               & $\mathcal{G}(0.096, 0.090)$       & $0.125 \pm 0.025$               \\
$u_2$                         & $\mathcal{G}(0.130, 0.065)$       & $0.126 \pm 0.031$               & $\mathcal{G}(0.252, 0.115)$       & $0.180 \pm 0.041$               \\
$v$ (day$^{-1}$)              & $\mathcal{U}(-0.1, 0.1)$          & $-0.000913 \pm 3.9\cdot10^{-5}$     & same as NIRSpec                   & $-0.000329 \pm 7.1\cdot10^{-5}$     \\
$c$                       & $\mathcal{U}(0.99, 1.01)$         & $1.0000976 \pm 5.4\cdot10^{-6} $   & same as NIRSpec                   & $1.0000339 \pm 8.2\cdot10^{-6}$   \\
$\sigma_s$ (ppm)               & $\mathcal{U}(0, 10^6)$            & $73.0 \pm 2.2$                    & same as NIRSpec                   & $112.0 \pm 3.0$\\ 
\hline
\end{tabular}
}
\end{table*}

\subsubsection{Light curve fitting}
After the third Eureka! stage, we used a custom Markov chain Monte Carlo-based (MCMC) light curve-fitting procedure. 
We used the \texttt{batman}-package for our forward model \citep{kreidberg_batman_2015}, adding together two transit models $T_i(t)$ combined with a linear slope in time to get our full transit model: 
\begin{equation}
\frac{F(t)}{F_*} = \left[c + v(t-t_{\text{Start}})\right]\left[T_b(t) + T_d(t)-1\right] \ .
\end{equation}

First, a broadband fit was performed to determine the orbital parameters of both TOI-270 d and b.
For this, we simply summed and then normalized the measured flux from both NRS1 and NRS2. Through a $4\sigma$ outlier rejection over a running mean after we rejected 20 of the remaining 1763 broadband data points.
The resulting light curve was then fit using the \texttt{emcee}-package \citep{Foreman_emcee_2013}, using 64 walkers, 20\,000 steps and a burn-in fraction of 40\%.
We assumed a circular orbit and fixed the orbital periods to previously derived values from \citet{vaneylen_toi270_2021} and used their constraints on the inclination as a Gaussian prior. 
In addition to the orbital parameters, we also fit a scatter parameter $\sigma_s$ for the whole light curve that inflates the error bars of each data point in the likelihood evaluation by 
\begin{equation}
\tilde\sigma_i=\sqrt{\sigma_i^2 + \sigma_s^2} \ .
\end{equation}

In light of recent work by \citet{coulombe_ld_2024} and \citet{keers_ld_2024}, we tested different limb-darkening approaches (see Appendix \ref{app:limb_darkening}). Based on that analysis, we used a quadratic limb-darkening law with a five-times widened Gaussian prior based on the ATLAS9 stellar model by \citet{castelli_atlas9_2004} generated with ExoCTK \citep{bourque_exoctk_2021}, using the stellar parameters from \citet{vaneylen_toi270_2021}.
In total, our NIRSpec broadband fit contained 13 parameters. The priors and median fit values are listed in Table \ref{table:broadband_results}, while the broadband fit itself is shown in
Fig. \ref{fig:broadband_lcs}.

For the spectral light curve fitting, we then fixed the orbital parameters to the median values derived from the broadband light curve. This resulted in seven free parameters for the fit: the two planetary radii, the two parameters each for limb-darkening and the linear trend, as well as the scatter value. The reduced number of free parameters allowed us to decrease the chain length to 10\,000 for the spectral light curve fitting, keeping all other settings the same.

The light curves were fit at the native resolution of NIRSpec. This has been shown to result in increased precision when binned afterwards and compared to data that was binned before fitting \citep{coulombe_ld_2024}. The limb-darkening priors were the only ones that differed from the broadband fit. The wavelength-specific priors were interpolated from the ATLAS9 model values.

\subsection{NIRISS observation}
Another transit of TOI-270 d was observed on 6 February 2024 in the NIRISS SOSS-mode as part of GO4098.
The observation contains 189 integrations of three groups with the SUBSTRIP256 readout pattern for a total exposure time of 4.0 hours within the 5.3 hour observation.

We extracted a spectrum from the first order of NIRISS that continually covers the $0.85-2.86\micron$ range, providing a significant improvement over the 28 Hubble WFC3 photometry data points around $1.4\micron$ \citep{mikalevans_toi270_2023} used by \citet{holmberg_possible_2024}. 
In principle, NIRISS also provides data at shorter wavelengths from its second order. However, we excluded the second-order NIRISS data from our analysis since we found evidence of strong stellar contamination (see Appendix \ref{app:niriss_order2}), making its interpretation challenging.

\subsubsection{Eureka! reduction}
With the goal of keeping both data reductions as similar as possible, we made use of Eureka!'s newly added support for NIRISS observations. 
Following our previous NIRSpec reduction, the .ecf-files are also available on zenodo.\footnote{\url{https://doi.org/10.5281/zenodo.15228454}} 
Our NIRISS Stages 1 and 2 included the same steps as the NIRSpec reduction, with some minor changes in the selected parameters. For the first stage, we also opted to skip the group-level background subtraction. 

Eureka! uses \texttt{pastasoss} \citep{pastasoss2_2023} to trace the spectral order and perform wavelength calibration, accounting for the minor offset that appears when the GR700XD grism is rotated into the optical path. In Stage 3, the pixels were shifted in accordance with the \texttt{pastasoss}-trace in each column to prepare for the spectral extraction.

A median background-subtraction was calculated on all pixels at a distance of more than 28 pixels away from the center of the trace and that do not belong to the other NIRISS orders. This removed potential $3\sigma$-outliers and produced background corrections of up to 215 $e^-/$s. As a last step, the spectrum was extracted with the same optimal spectrum extraction algorithm, using a half-width of 21 pixels.

\subsubsection{Light curve fitting}
After Stage 3 of the Eureka! pipeline, we proceed analogously to the NIRSpec data, by fitting the broadband light curve using MCMC. 

The employed model, however, is more simplified compared to the latter since it only had to account for a single planet. All other MCMC parameters and priors are identical to those from the NIRSpec reduction. One minor difference between the two reductions is the skip of the outlier-removal step we performed for NIRSpec. The median values for the NIRISS observation are listed in Table \ref{table:broadband_results} while the model and fit residuals are shown in Fig. \ref{fig:broadband_lcs}.

\begin{figure*}[ht]
  \centering
  \resizebox{\hsize}{!}{\includegraphics{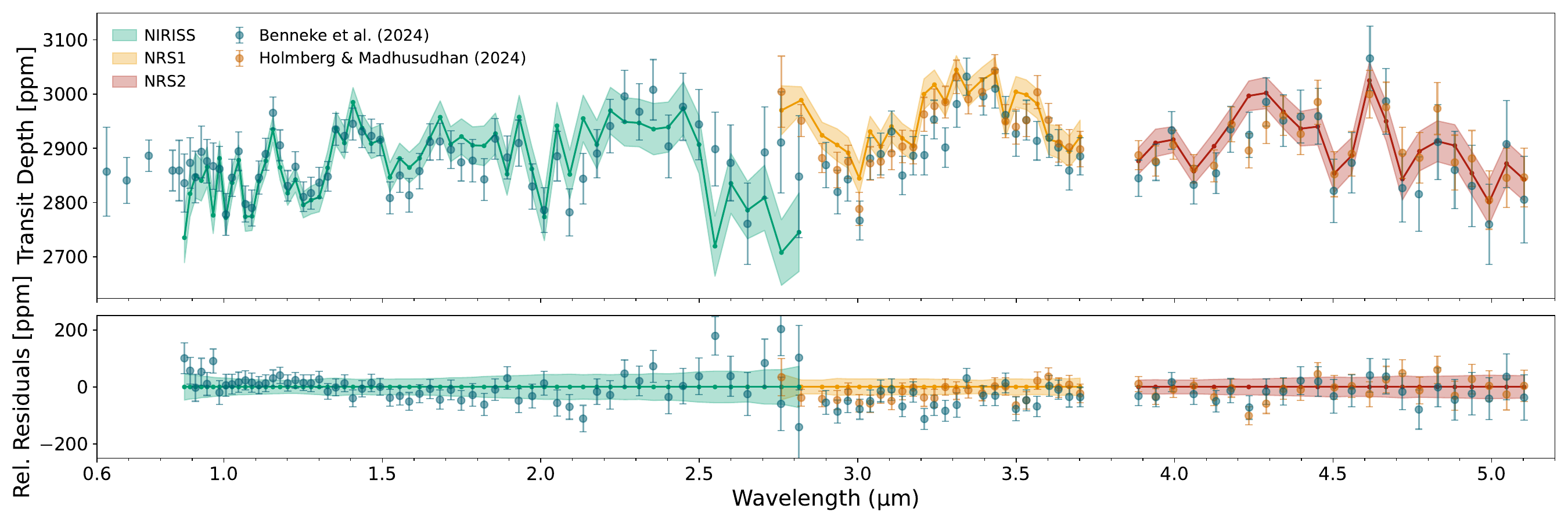}}
  \caption{Transit spectrum of TOI-270 d binned to R $\approx50$ for comparison to \citet{benneke_jwst_2024} and \citet{holmberg_possible_2024}. We note that the spectrum from this work and \citet{holmberg_possible_2024} were originally reduced and fit at the native instrument resolution. We see excellent agreement over most of the probed wavelength range, with only the $\geq2.5\micron$ portion of our NIRISS spectrum deviating significantly from the other reduction and diverging from the shortest wavelength NRS1 data. We test for the impact of these data in \ref{app:niriss_beyond2.5}. In general, our uncertainties are slightly larger than in \citet{holmberg_possible_2024} but significantly smaller for the NIRSpec data when compared to \citet{benneke_jwst_2024}. No offsets have been applied to the spectra.}
  \label{fig:transit_spectrum}%
\end{figure*}
For the spectral light curve fitting, the orbital parameters are again fixed to the parameters from the broadband fit. This results in a model with six free parameters that were each fit at the native NIRISS resolution of the first-order trace. The resulting transmission spectrum for TOI-270 d is shown in Fig. \ref{fig:transit_spectrum}.

\subsection{TOI-270 d transmission spectrum}

The transmission spectrum of TOI-270 d obtained by the NIRISS and the NIRSpec instruments is shown in Fig. \ref{fig:transit_spectrum} and the corresponding Allan-variance can be found in Appendix \ref{app:lc_diagnostics}. 
Our derived broadband parameters do largely agree with the reported parameters from \citet{benneke_jwst_2024} and \citet{holmberg_possible_2024}. 
With $96\,\text{ppm}$, our NIRSpec broadband light curve exhibits similar scatter than the $99-100\,\text{ppm}$ obtained by \citet{holmberg_possible_2024}. This is likely caused by a combination of the different background $\sigma$-threshold, applied bias correction and the more restrictive jump-detection threshold.

Our NIRSpec data reduction and the one by \citet{holmberg_possible_2024} were both performed at the native resolution of the instrument, while \citet{benneke_jwst_2024} performed a data binning before the reduction. Nonetheless, as Fig. \ref{fig:transit_spectrum} suggests, the \citet{holmberg_possible_2024} and our reduction when binned down to the resolution of the \citet{benneke_jwst_2024} data, seem to agree quite well. This indicates that the two different approaches -- binning before and after the data reduction of the spectral light curve -- seem to yield roughly the same result in this case. It is, however, noteworthy that the spectra binned after the data reduction generally have smaller uncertainties, as predicted in \citet{coulombe_ld_2024}.

The only part of the spectrum where significant disagreement can be noticed is the red end of the NIRISS first-order spectrum. Here we obtain transit depths that are up to $100\,\text{ppm}$ below the values from \citet{benneke_jwst_2024} and also in disagreement with our own NIRSpec NRS1 data. An investigation on the impact of these differences can be found in Appendix \ref{app:niriss_beyond2.5}. However, we find that this part of the spectrum is not critical for our results and is an outcome of our native resolution analysis.

\section{Atmospheric retrieval calculations}
\label{chap:retrieval}

For the atmospheric characterization of TOI-270 d, we employed the open-source Bern Atmospheric Retrieval code (\textsc{BeAR})\footnote{\url{https://github.com/NewStrangeWorlds/BeAR}}, a renamed version of the former \textsc{Helios-r2} code \citep{kitzmann_heliosr2_2020}. If not mentioned otherwise, we assumed a cloud-free atmosphere with an isothermal temperature $T_\mathrm{p}$, vertically constant volume mixing ratios $x_i$ of chemical species, and a constant surface gravity $g$\footnote{We use cgs units for the surface gravity $g$ throughout this study, unless stated otherwise.}. The atmosphere was divided into 100 levels, i.e. 99 layers, equally spaced in $\log(p)$ from $10$ bar to $10^{-8}$ bar. The planet radius $R_\mathrm{p}$ used in our retrieval calculations corresponds to the radius at the bottom of our assumed atmosphere.  

Our high-resolution forward model spectra were generated at a spectral resolution of $R=\lambda/\Delta\lambda$ of 10\,000 and then binned down to the corresponding JWST instrument resolution. We have tested the impact of using resolutions as high as $R = 100\,000$ but did not find deviations from the lower resolution spectra. Due to much lower computational demand, we, therefore, used $R = 10\,000$ throughout this study, unless stated otherwise.

Since the atmospheric retrievals were focused on our reduced NIRSpec and NIRISS data, we also allowed for an offset between the NRS1 and NRS2 ($\Delta_\mathrm{NRS2}$) detectors of NIRSpec, as well as between NIRSpec NRS1 and NIRISS ($\Delta_\mathrm{NIRISS}$).

\textsc{BeAR} uses a Bayesian nested sampling method \citep{Skilling2004AIPC..735..395S} to obtain the posterior distributions of the free parameters and the Bayesian evidence $\mathcal Z$. The Bayesian evidence can also be used to perform model comparisons. For two different models $i$ and $j$, one can compute the so-called Bayes factor,
\begin{equation}
  B_{ij} = \frac{\mathcal Z_i}{\mathcal Z_j} \ ,
\end{equation}
quantifying the strength of evidence in favor of model $i$ over $j$ to represent the measured data. On the Jeffreys scale \citep{doi:10.1080/01621459.1995.10476572}, a value of 1, 3.2, and 10 correspond to no, substantial, and strong evidence in favor of model $i$ over $j$, respectively. A decisive evidence is categorized by $B_{ij} > 100$.
Often, this is converted into a frequentist representation in terms of standard deviations

The posterior calculations are performed through the \texttt{MultiNest} library \citep{multinest1, multinest2, multinest3}, a widely used implementation of nested sampling. In our calculations, we used \texttt{MultiNest} with 800 live points. Table \ref{table:retrieval_priors} shows the prior distributions for the various free parameters used in our retrieval model.

\begin{table}[h]
\begin{center}
\renewcommand*{\arraystretch}{1.3}
\caption{Prior distributions for all parameters used for the mentioned atmospheric retrieval cases of TOI-270 d.
\label{table:retrieval_priors}}
\begin{tabular}{ll}
\hline \hline
Parameter                                & Prior\\ 
\hline
$R_*$ $(R_\odot)$                        & $\mathcal{G}(0.378, 0.011)\,^*$\\
$R_{\mathrm{p}}$ $(\mathrm{R}_{\oplus})$ & $\mathcal{U}(1.8, 2.3)$\\
$T_\mathrm{p}$ (K)                       & $\mathcal{U}(100, 800)$\\
$\log{g}$                                & $\mathcal{G}(3.0126, 0.046)\,^*$\\
$x_i$                                    & $\mathcal{LU}(10^{-12}, 10^0)$\\
$\Delta_\mathrm{NIRISS}$ (ppm)           & $\mathcal{U}(-100, 100)$\\
$\Delta_\mathrm{NRS2}$ (ppm)             & $\mathcal{U}(-100, 100)$\\
\hline
\end{tabular}
\end{center}
\tablebib{(*) Taken or calculated from \citet{vaneylen_toi270_2021}.
}
\end{table}

\subsection{Chemical species}

Our fiducial model contained the following molecules: methane (\ch{CH4}), carbon dioxide (\ch{CO2}), water (\ch{H2O}), carbon monoxide (CO), ammonia (\ch{NH3}), carbon disulfide (\ch{CS2}), and sulfur dioxide (\ch{SO2}). This selection corresponds to same set of species that has been previously analyzed by \citet{benneke_jwst_2024}. The background atmosphere is assumed to be filled by a mixture of molecular hydrogen (\ch{H2}) and helium (He), assuming a solar H/He element abundance ratio. The mean molecular weight $\bar\mu$ is calculated from the abundances of the chemical species included in the retrieval.

In addition to these chemical species, we also performed additional retrieval calculations testing for the presence of various other molecules. This includes methyl chloride (\ch{CH3Cl}), methyl fluoride (\ch{CH3F}), carbonyl sulfide (\ch{OCS}), hydrogen sulfide (\ch{H2S}), thioformaldehyde (\ch{H2CS}), and a broad range of different sulfur-bearing species. These species are described in more detail in Sect. \ref{sec:retrieval_add_molecules} about our different retrieval scenarios.

\subsection{Opacity data}

The opacity data has been obtained by using the open-source code \textsc{HELIOS-K} \citep{Grimm2015ApJ...808..182G, grimm_heliosk_2021}\footnote{\url{https://github.com/exoclime/HELIOS-K}}. The corresponding line lists used to calculate the absorption coefficients are listed in Table \ref{table:linelists} in Appendix \ref{sec:app_opacities} and were mostly taken from the ExoMol database \citep{Tennyson2024JQSRT.32609083T}. Some opacities are based on the line lists available in HITRAN \citep{hitran_2020}. The calculated high-resolution opacity grids for the species used in our retrieval are available on the DACE platform\footnote{\url{https://dace.unige.ch/}}. The only exception is dimethyl sulfide (\ch{(CH3)2S}, DMS), where line-list data is not available. Here, we used measured cross-sections published in the HITRAN database. We note that this data is very limited in terms of temperature and pressure information, as well as spectral resolution. Only three temperatures below 330 K and a single pressure of 1 bar are available. However, the 1 bar is mostly composed of a molecular nitrogen (\ch{N2}) background gas.

Besides these molecular opacities, we also included collision induced absorption (CIA) by \ch{H2}-He and \ch{H2}-\ch{H2} pairs from the HITRAN database. In addition to the absorption cross-sections described above, we also added  Rayleigh scattering by \ch{H2} and He as opacity sources, which could potentially be important at the blue end of the NIRISS data.

\subsection{Line spread function}

We performed the retrieval calculations at the native resolution of the NIRISS and NIRSpec G395H instruments. At this resolution level, the line spread function (LSF) becomes important to simulate the spectrum at the pixel level. 

The LSF describes the distribution of light as a function of wavelength on the detector in response to a monochromatic source. It causes light of a discrete wavelength to be spread across several pixels, which is usually described by a Gaussian line profile with a certain full-width at half-maximum (FWHM) \citep[e.g.][]{moore_IPC-effect_2004}. 

\textsc{BeAR} takes this effect into account by performing a convolution of the line profile with the computed high-resolution spectrum before binning it down to the instrument resolution. The integration is limited to a distance of 5 $\sigma$ from the center wavelength to limit the computational costs of the convolution.

We approximated the wavelength-dependent FWHM of the LSF by $FWHM=\lambda/R(\lambda)$ where $R =\lambda/ \Delta \lambda$ corresponds to the spectral resolution of the instrument. The spectral resolving power was obtained from data provided in \citet{pontoppidan_pandeia_2016}. For the first two NIRISS spectral orders and NIRSpec NRS1 and NRS2, this yields a FWHM of roughly two pixels across the spectrum.

\section{Retrieval results}
\label{section:retrieval_results}

In this section we present the results of our atmospheric retrieval calculations. We first show the ones for our fiducial model and test more complex models afterwards. Finally, we also discuss the robustness of the results on the potential binning of the observational data and compare our retrieval framework with results from \citet{benneke_jwst_2024} and \citet{holmberg_possible_2024}.

\subsection{Fiducial model results}

\begin{figure*}[t!]
  \centering
  \resizebox{\hsize}{!}{\includegraphics{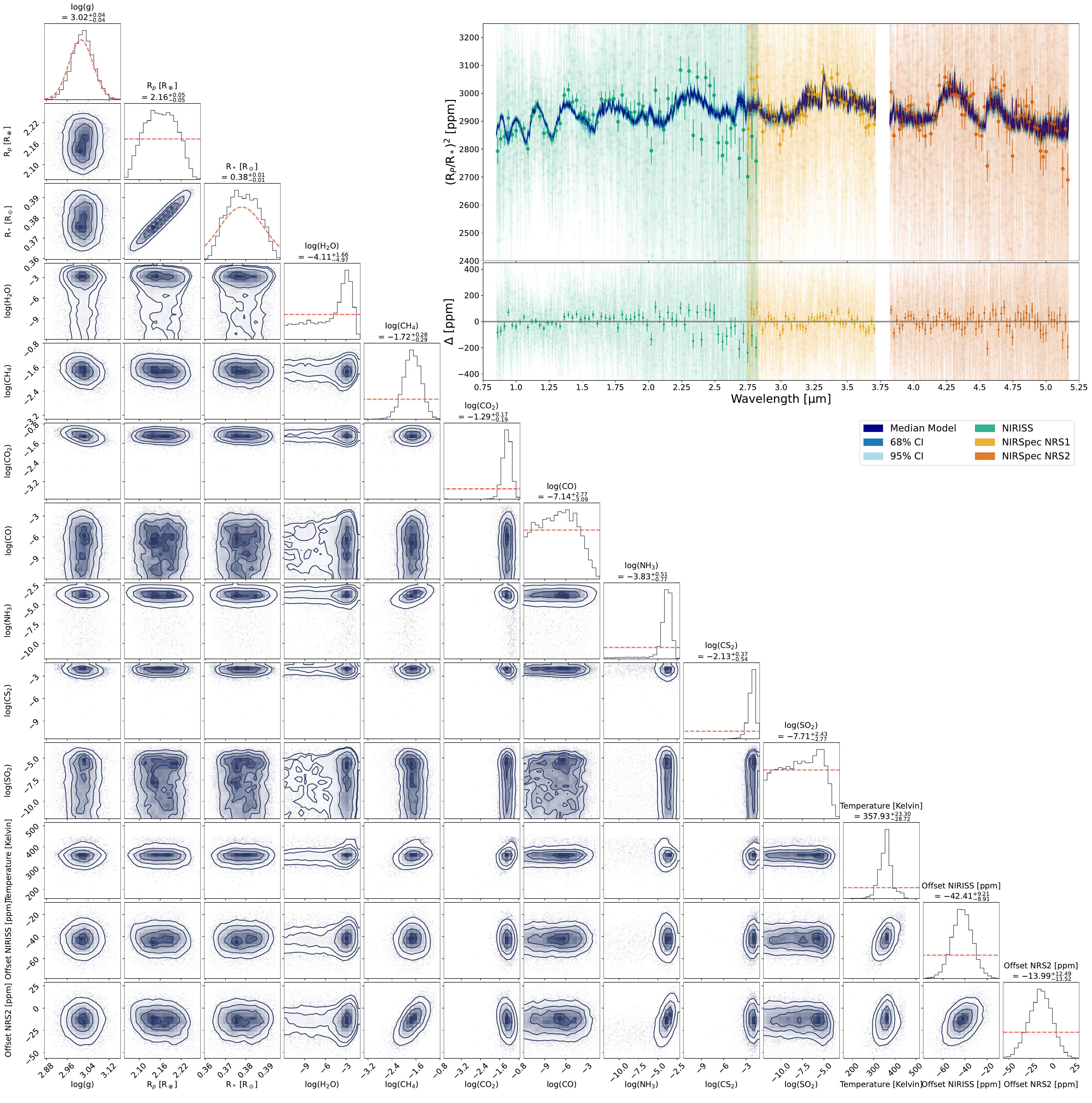}}
  \caption{Posterior distributions and transit spectra of our fiducial model at the full spectral resolution of NIRISS SOSS GR700XD and NIRSpec G395H. The prior distributions for the retrieval parameters are shown in red. The top right shows the resulting transit spectrum models and the initial native resolution data. For visual aid, we also overplot a binned set of points for each dataset, the 32-pixel binning.}
  \label{fig:corner_defaultmodel}%
\end{figure*}

The atmospheric retrieval posterior distribution for our fiducial model calculated with \textsc{BeAR} are shown in Fig. \ref{fig:corner_defaultmodel}, together with the median spectrum and its confidence interval from the posterior sample. The posterior values are summarized in Table \ref{table:post_defaultmodel}. Besides the directly retrieved parameters, the table also contains derived ones, such as the C/O ratio, various metallicities ($[\mathrm{M/H}]$, $[\mathrm{S/H}]$, $Z$), or the planet's mass. 

The Bayes factor for each molecule of this model are listed in the first row of Table \ref{table:bayes_factors}. The detection significance is calculated by running nested sampling retrievals on the same dataset, leaving out one species at a time and comparing its Bayesian evidence to the one of our fiducial model via the Bayes factor \citep[e.g.][]{Benneke_bayesfactor_2013, Trotta_bayesinthesky_2008}.

\begin{table*}[t]
\begin{center}
\renewcommand*{\arraystretch}{1.3}
\caption{Median values and $1\sigma$-intervals from the posterior distributions for the retrieval calculations of TOI-270 d described in the text.
}
\label{table:post_defaultmodel}
\begin{tabular}{lccccccc}
\hline \hline
Parameter                & \multicolumn{7}{c}{Model}             \\ 
                         & Fiducial                & \ch{CH3Cl}        & \ch{CH3F}         & Sulfur           & Sulfur + DMS  & Sulfur + \ch{CH3Cl} & Sulfur + \ch{CH3F} \\
\hline
$\ln \mathcal Z$         & -38406.30               & -38394.70         & -38396.56         & -38397.82        & -38398.92              & -38397.37         & -38394.87             \\
\textbf{Retrieved}       &                         &                   &                   &                   &                   &                       &\\
$R_\mathrm{p}$ (\Rearth) & $2.16 \pm 0.05$         & $2.16 \pm 0.05$   & $2.16 \pm 0.05$         & $2.17 \pm 0.05$         & $2.16 \pm 0.05$         & $2.16 \pm 0.05$         & $2.16 \pm 0.05$ \\
$T_\mathrm{p}$ (K)       & $357^{+23}_{-29}$       & $364^{+45}_{-48}$ & $342^{+40}_{-46}$       & $371^{+48}_{-46}$       & $374^{+46}_{-44}$       & $378^{+45}_{-44}$       & $350^{+40}_{-43}$ \\
$\log x_{\ch{H2O}}$      & $-4.11^{+1.66}_{-4.97}$ & $-2.51^{+0.72}_{-1.20}$ & $-2.54^{+0.74}_{-1.4}$  & $-3.22^{+0.66}_{-0.83}$ & $-3.16^{+0.66}_{-0.85}$ & $-2.91^{+0.65}_{-0.82}$ & $-2.92^{+0.63}_{-0.73}$ \\
$\log x_{\ch{CH4}}$      & $-1.72^{+0.28}_{-0.29}$ & $-2.11^{+0.37}_{-0.40}$ & $-2.00^{+0.32}_{-0.39}$ & $-2.61^{+0.36}_{-0.38}$ & $-2.59^{+0.39}_{-0.38}$ & $-2.60^{+0.39}_{-0.39}$ & $-2.58^{+0.37}_{-0.39}$ \\
$\log x_{\ch{CO2}}$      & $-1.29^{+0.17}_{-0.19}$ & $-1.35^{+0.19}_{-0.27}$ & $-1.38^{+0.19}_{-0.24}$ & $-1.90^{+0.36}_{-0.46}$ & $-1.94^{+0.37}_{-0.46}$ & $-1.90^{+0.37}_{-0.47}$ & $-1.85^{+0.34}_{-0.45}$ \\
$\log x_{\ch{CO}}$       & $<-4$                   & -                 & -                       & -                       & -                       & -                             & -\\
$\log x_{\ch{NH3}}$      & $-3.83^{+0.51}_{-0.77}$ & $-4.14^{+0.78}_{-2.80}$ & $-3.80^{+0.65}_{-1.60}$ & $-5.9^{+1.6}_{-3.9}$   & $-5.2^{+1.1}_{-3.0}$  & $-4.84^{+0.94}_{-3.60}$ & $-4.61^{+0.81}_{-2.40}$ \\
$\log x_{\ch{CS2}}$      & $-2.13^{+0.37}_{-0.54}$ & $-1.98^{+0.40}_{-0.56}$ & $-2.20^{+0.42}_{-0.60}$ & $-2.56^{+0.49}_{-0.59}$ & $-2.60^{+0.51}_{-0.60}$ & $-2.50^{+0.50}_{-0.62}$ & $-2.65^{+0.51}_{-0.59}$ \\
$\log x_{\ch{SO2}}$      & $< -5$                  & -                 & -                       & -                       & -                       & -                           & -\\ 
$\log x_{\ch{CH3Cl}}$    & -                       & $-3.17^{+0.93}_{-0.43}$                     & -                       & -                       & - & $-3.73^{+0.51}_{-0.93}$     & -                   \\ 
$\log x_{\ch{CH3F}}$     & -                       & -                 & $-3.20^{+0.35}_{-0.40}$ & -                       & - & -                       & $-3.68^{+0.40}_{-0.43}$       \\ 
$\log x_{\ch{(CH3)2S}}$  & -                       & -                 & - & -                       & $-6.20^{+1.40}_{-3.50}$ & -                       & -                             \\ 
$\log x_{\ch{H2CS}}$     & -                       & -                 & -                       & $-3.68^{+0.47}_{-0.56}$ & $-3.97^{+0.60}_{-1.10}$ & $< -4$                    & $< -4$                      \\ 
$\log x_{\ch{CS}}$       & -                       & -                 & -                       & $-1.24^{+0.18}_{-0.27}$ & $-1.23^{+0.18}_{-0.26}$ & $-1.25^{+0.18}_{-0.28}$ & $-1.34^{+0.19}_{-0.30}$       \\
$\Delta_\mathrm{NIRISS}$ (ppm) & $-42.4 \pm 9$ & $-29.0 \pm 11$ & $-28.7 \pm 12$       & $-31.1 \pm 11$      & $-28.0 \pm 11$  & $-25.8 \pm 11$       & $-23.8 \pm 11$                                \\
$\Delta_\mathrm{NRS2}$ (ppm)   & $-14.0 \pm 13$ & $-17.7 \pm 12$ & $-14.1 \pm 14$       & $-33.5 \pm 13$       & $-28.8 \pm 14$ & $-25.3 \pm 13$       & $-24.8 \pm 14$                               \\
\hline
\textbf{Derived}         &  & & & & & &\\
$\bar{\mu}$ (amu)        & $5.56^{+1.10}_{-0.84}$ & $5.57^{+1.10}_{-0.87}$ & $5.08^{+0.95}_{-0.84}$ & $5.83^{+1.20}_{-0.95}$ & $5.77^{+1.10}_{-0.92}$ & $5.88^{+1.10}_{-0.92}$ & $5.35^{+0.96}_{-0.83}$ \\
C/O                      & $0.77^{+0.28}_{-0.14}$ & $0.68^{+0.36}_{-0.12}$ & $0.68^{+0.28}_{-0.12}$ & $2.95^{+5.7}_{-1.7}$ & $3.21^{+6.1}_{-1.9}$ & $2.83^{+5.5}_{-1.6}$ & $2.18^{+3.8}_{-1.1}$ \\
$[\mathrm{M/H}]$         & $2.20 \pm 0.14$ & $2.17 \pm 0.15$ & $2.12^{+0.14}_{-0.17}$ & $2.14 \pm 0.13$ & $2.09 \pm 0.14$ & $2.15 \pm 0.14$ & $2.14 \pm 0.14$ \\
$[\mathrm{S/H}]$         & $2.85^{+0.37}_{-0.54}$ & $3.00^{+0.39}_{-0.56}$ & $2.78^{+0.42}_{-0.60}$ & $3.51^{+0.16}_{-0.20}$ & $3.51^{+0.16}_{-0.21}$ & $3.51^{+0.16}_{-0.21}$ & $3.40^{+0.18}_{-0.24}$ \\
$Z$                      & $0.61 \pm 0.08$ & $0.61 \pm 0.08$ & $0.57 \pm 0.09$ & $0.63 \pm 0.07$ & $0.63 \pm 0.08$ & $0.64 \pm 0.08$ & $0.60 \pm 0.08$ \\
$M_\mathrm{p}$ ($\mathrm{M}_\oplus$)  & $4.97^{+0.56}_{-0.48}$ &$4.99^{+0.56}_{-0.52}$ &$4.96^{+0.58}_{-0.50}$ &$4.86^{+0.53}_{-0.46}$ & $4.88^{+0.50}_{-0.47}$ &$4.87^{+0.55}_{-0.47}$ &$4.82^{+0.55}_{-0.47}$ \\ 
\hline
\end{tabular}
\end{center}
\end{table*}

The results suggests an atmosphere with a mean molecular weight of 5.56 amu, almost twice of a solar-elemental abundance composition. This increased value of $\bar \mu$ is mainly driven by a high abundance of carbon-bearing molecules, such as \ch{CH4}, \ch{CO2} and \ch{CS2}. All three combined constitute about 8\% of the planet's atmosphere. This is also reflected in the derived overall metallicity $[\mathrm{M/H}]$, which is almost 160 times the solar value. Following the Bayes factors shown in Table \ref{table:bayes_factors}, models with methane and carbon monoxide are overwhelmingly preferred, while the significance for \ch{CS2} is moderate. The only carbon-bearing species that has  an unconstrained mixing ratio is carbon monoxide. Here, we only obtain an  upper limit of around $10^{-4}$.

Water is constrained at a relatively low abundance of only $10^{-4}$, which is surprising given the increased atmospheric metallicity. The retrieved temperature of the atmosphere of about 360 K suggests that water should not be condensed yet, at least in the part of the modeled atmosphere down to a maximum pressure of 10 bar. However, the Bayes factor for \ch{H2O} is only around unity, suggesting that the retrieval does not prefer to include the molecule. This essentially non-detection of water could suggest that it is condensed out somewhere below the part of the atmosphere that is accessible to transmission spectroscopy.

We constrain the abundance of both methane and carbon dioxide, with median mixing ratios of about $10^{-2.1}$ and $10^{-1.3}$. The occurrence of both molecules at such low temperatures and especially the very high mixing ratio of \ch{CO2} mirror the results obtained by \citet{madhusudhan_carbon_2023} for the even colder sub-Neptune K2-18 b. For ammonia we obtain an abundance of about $10^{-3.8}$. However, as shown in Table \ref{table:bayes_factors}, the Bayes factor of the model that includes \ch{NH3} is only around 4, which indicates a weak detection.

Based on the constrained abundances of the species included in the retrieval we also derived the C/O ratio. With a value of about $0.77$ it is super-solar, but still oxygen-dominated. If water would indeed be condensed in the lower atmosphere, the C/O ratio of the bulk composition would likely be smaller and closer to the solar value of 0.55. 

One surprising outcome is the very high abundances of $10^{-2.1}$ for the sulfur-bearing species \ch{CS2}. The other sulfur species included in the retrieval, \ch{SO2}, only yields an upper limit for its mixing ratio. The latter model also has a Bayes factor of less than one, indicating that a model without this molecule is preferred. The strong presence of carbon disulfide is also reflected in the $[\mathrm{S/H}]$ element abundance ratio, which is more than 700 times its solar value and also much larger than the general $[\mathrm{M/H}]$. This suggests that the atmosphere could be strongly enriched in sulfur. 

\begin{figure*}[t!]
  \centering
  \resizebox{\hsize}{!}{\includegraphics{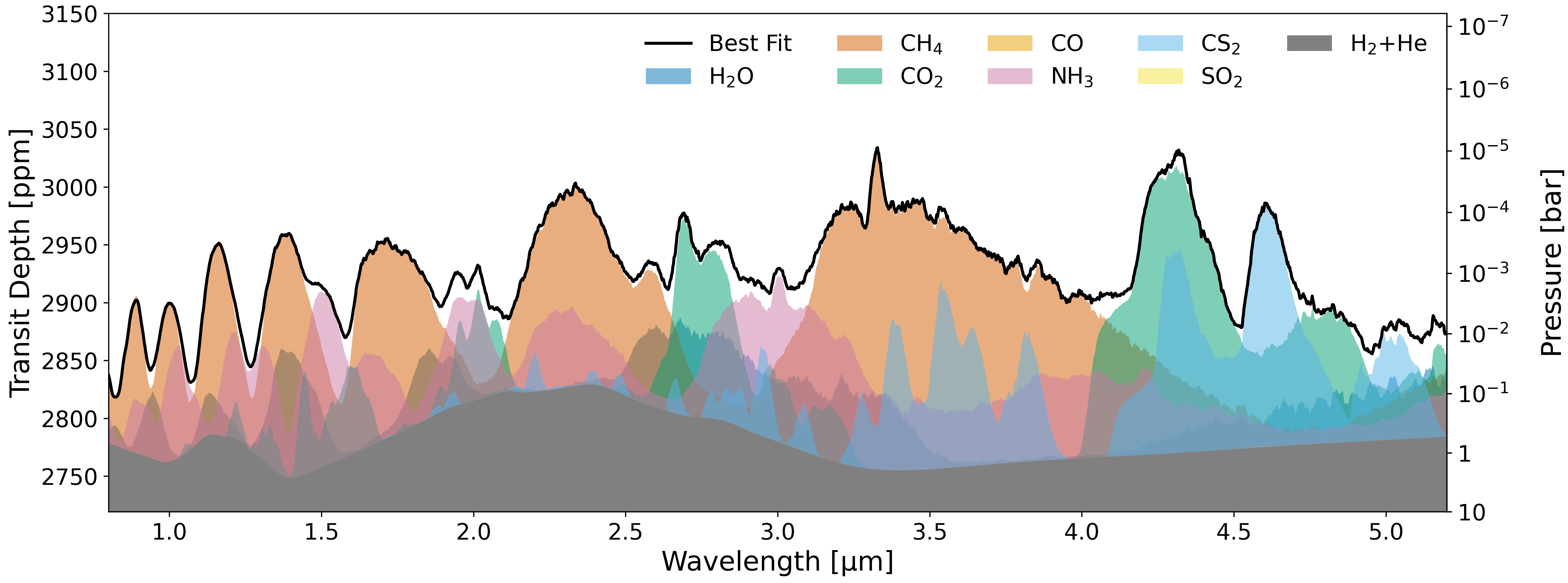}}
  \caption{Breakdown of our best-fit model for the transit spectrum of TOI-270 d. Absorption bands of CH$_4$, CO$_2$ are clearly visible and one isolated absorption band of CS$_2$ contributes at $4.6\micron$. The impact of water, CO and SO$_2$ on the spectrum is negligible. The model has been smoothed for visual clarity.}
  \label{fig:bestfit_mols}
\end{figure*}

In addition to the posteriors, we also show the contribution of each molecule to the best-fit spectrum of the posterior sample in Fig. \ref{fig:bestfit_mols}. As the figure suggests, strong absorption bands of \ch{CH4} and \ch{CO2} are visible in the spectrum, which explains the overwhelming preference of the models including these species. Water shows only very minor contributions to the overall spectrum due to its very low abundance. The weakly-preferred \ch{NH3} only has a very few wavelength regions where its contribution can be clearly identified. 

In terms of species detections, our results broadly agree with those by \citet{holmberg_possible_2024} and \citet{benneke_jwst_2024}. Both also found evidence for \ch{CH4} and \ch{CO2}, as well as \ch{H2O} and \ch{CS2}. Our abundance constraints for methane and carbon dioxide are very similar to the one by \citet{benneke_jwst_2024} with the one important difference that using our data reduction yields a higher abundance of \ch{CO2} than \ch{CH4}. The abundances of these two species, however, differ by roughly one order of magnitude compared to those presented by \citet{holmberg_possible_2024}. Neither of these studies found evidence for \ch{CO} which is also the outcome of our retrieval calculations. We do also not detect \ch{SO2}, confirming the results by \citet{holmberg_possible_2024}, while \citet{benneke_jwst_2024} claimed to see signs for this species, but did not present a clear detection.

One notable difference between our results and those by \citet{holmberg_possible_2024} is the atmospheric temperature. The latter study obtained significantly lower temperatures of 300 K and less but also employed a more complex temperature profile. Our isothermal temperature of about 360 K, is much closer to the value of roughly 385 K obtained by \citet{benneke_jwst_2024}. The impact of a more complex temperature profile on our retrieval results is tested in Sect. \ref{sec:non_iso_tp}.

Our derived planet mass of about $4.97 \pm 0.56 \,\text{M}_\oplus$ is consistent with the radial-velocity measurement published by \citep{vaneylen_toi270_2021}. In Appendix \ref{chap:mass_comparison} we additionally test the mass constraints from the two previous studies by \citet{benneke_jwst_2024} and \citet{holmberg_possible_2024} for consistency. We find that only the retrieved chemistry of the former is consistent with the measured planet mass without the addition of a weakly absorbing background species. One can enforce mass-consistency also for the composition in \citet{holmberg_possible_2024} through a strong prior on the surface gravity, but this yields an inconsistent stellar radius instead.

Finally, we do find a non-negligible offset of about -42 ppm between the NIRISS and NIRSpec NRS1 data. On the other hand, the offset between the two NIRSpec detectors is almost consistent with zero ($-14.0 \pm 13.5$ ppm). 

\subsection{Impact of data binning on retrieval calculations}

In this section we explore the impact of data binning on the outcome of the retrieval calculations. In the models discussed above, we used the observational data at its native resolution, together with a corresponding LSF. 

\begin{table*}[]
\caption{Calculated Bayesian Evidence of the default models at each resolution configuration and Bayes factor for all included molecules.
\label{table:bayes_factors}}
\centering
\begin{tabular}{lcccccccc}
\hline \hline
Data binning  & Fiducial model   & CH$_4$  & CO$_2$  & H$_2$O  & CS$_2$  & NH$_3$  & CO  & SO$_2$ \\ 
\cline{2-9}
\multicolumn{1}{c}{}  & $\log\mathcal{Z}$  & \multicolumn{7}{c}{Bayes factor with respect to  ($B_{\text{Def, i}}$)}                                                                                                                                                            \\ \hline
Full with LSF      & -38406.30    & $3.15\times10^{25}$ & $3.28\times 10^7$ & 1.15                     & 78.0 & 4.08 & 0.56                     & 0.40                     \\
Full w/o LSF                   & -38420.77 & $7.29\times 10^{19}$                    & $3.78\times 10^{3}$                  & 0.83 & 24.8     & 1.16     & 1.04 &2.40 \\
2-pixel                                                        & -18450.06                        & $6.11\times10^{22}$ & $2.83\times10^{5}$  & 0.64                     & 31.3 & 12.4 & 0.67                     & 0.92                   \\
4-pixel                                                        & -8852.38                         & $4.04\times10^{21}$ & $4.59\times10^4$  & 0.68                     & 11.9 & 3.32 & 0.81                     & 0.91                     \\
8-pixel                                                        & -4232.96                         & $3.12\times10^{23}$ & $2.34\times10^5$  & 0.82                        & 3.71 & 9.25 & 0.81                     & 0.52                  \\
16-pixel                                                       & -2037.11                         & $4.84\times10^{25}$ & $0.55\times 10^7$ & 1.64                     & 4.38 & 31.6 & 1.48                     & 0.67                     \\
32-pixel                                                       & -982.74                          & $7.56\times10^{25}$ & $6.39\times 10^7$ & 2.16                     & 3.40 & 41.4 & 0.94                     & 0.88                     \\ 
R $\approx50$                & -603.91                          & $3.68\times10^{25}$ & $1.35\times 10^{10}$ & 2.02                     & $3.13\times 10^{4}$ & 41.7 & 0.81                     & 0.65                    \\ \hline
\end{tabular}%
\end{table*}

\begin{figure*}[ht]
  \centering
  \resizebox{\hsize}{!}{\includegraphics{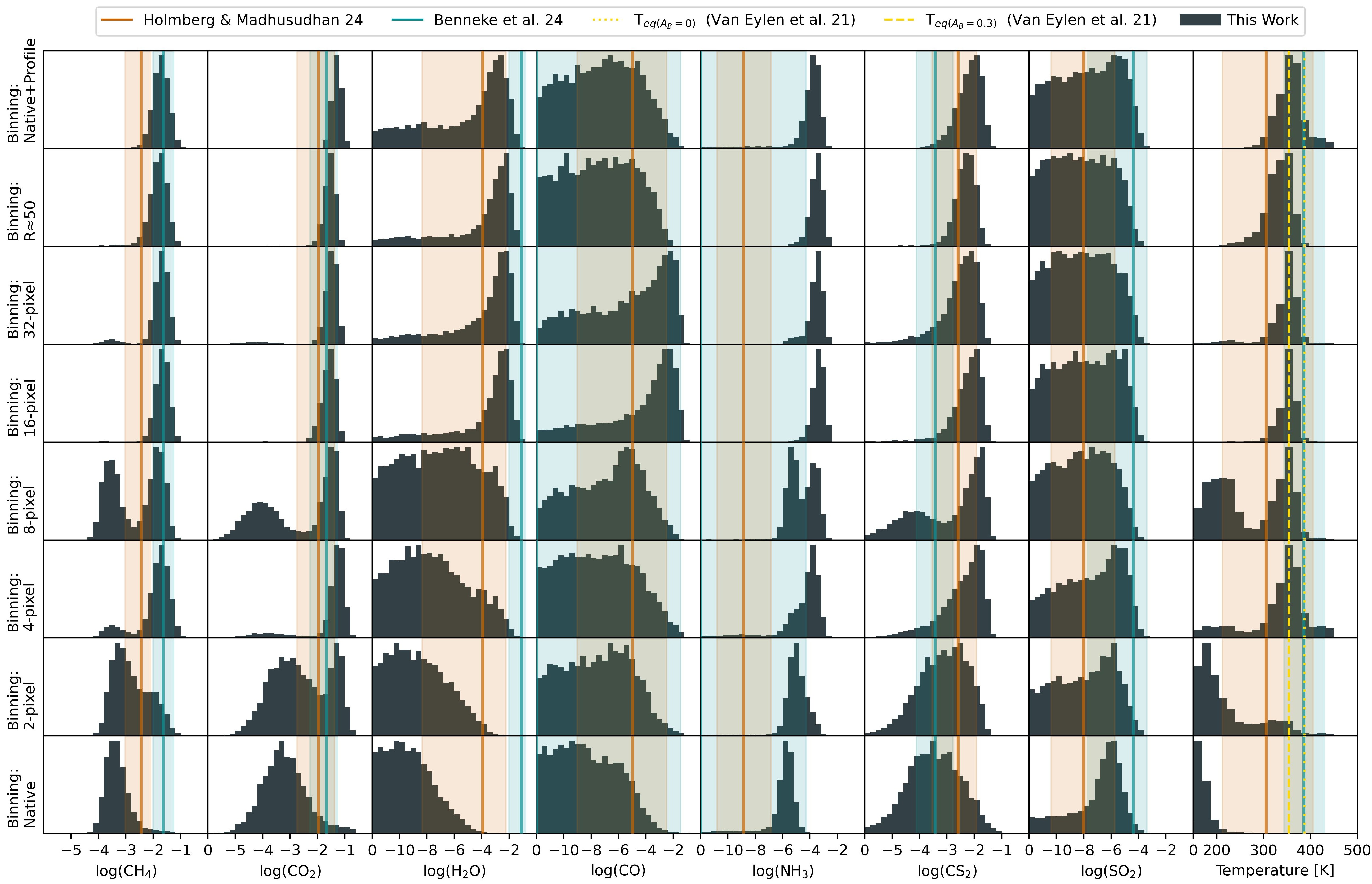}}
  \caption{Histograms of the posterior distributions for all chemical species and the temperature of our retrieval calculations listed in Table \ref{table:bayes_factors}. The corresponding retrieved median parameters and $1\sigma$ intervals from \citet{benneke_jwst_2024} and the canonical retrieval from \citet{holmberg_possible_2024} are shown for comparison.}
  \label{fig:retrieval_posterior_summary}%
\end{figure*}

Here, we performed additional retrieval calculations on data that has been binned over every 2, 4, 8, 16, and 32 pixels. These retrievals do not use the LSF. Additionally, we also test a data binning using the wavelength bins from \citet{benneke_jwst_2024} that correspond to a spectral resolution of around 50. The latter case has the fewest number of data points of all test cases. Finally, we also include a test case using the native resolution but without the LSF. The impact of the data binning on the detection of molecules is summarized in Table \ref{table:bayes_factors}, while the posterior distributions are shown in Fig. \ref{fig:retrieval_posterior_summary}.

Our fiducial, full resolution retrieval including the LSF is in good agreement with the parameters obtained using bin sizes of either 16 or 32 pixels, as well as the R $\approx50$ binning. At these rather coarse binnings, the impact of the LSF is largely decreased.

All other binnings using fewer pixels, however, are strongly affected. As the results shown in Fig. \ref{fig:retrieval_posterior_summary} indicate, the retrieval calculation often yield bimodal distributions in the abundances of the atmospheric species as well as atmospheric temperatures. Some of the latter converge to the lower edge of their prior. As a consequence of these lower temperatures, the overall atmospheric metallicity is decreasing, allowing the retrieval to compensate for the otherwise shrinking scale height.

With respect to the detection of molecules, Table \ref{table:bayes_factors} clearly indicates that the inclusion of CH$_4$ and CO$_2$ is favored across all resolutions. This suggests that molecules with broad and easily identifiable absorption bands in the spectrum can be recognized at either native or lower resolutions. However, as discussed above and shown in Fig. \ref{fig:retrieval_posterior_summary}, the actual posterior distributions can differ, with some of them showing a bimodal distribution. In contrast to that, the inclusion of \ch{H2O}, CO, and \ch{SO2} yield inconclusive Bayes factors for all considered data binnings. 

The most interesting cases are \ch{CS2} and \ch{NH3}, where the Bayes factor varies significantly over the range of data resolutions.
This is worrying, considering that the retrieved abundances are largely consistent between the native resolution using the LSF and a pixel binning over $\geq 16$ pixels. Since the prior distributions are identical in all cases and the posterior distributions are very similar (see Fig. \ref{fig:retrieval_posterior_summary}), the large difference in the Bayes factor is likely originating from the evaluations of the likelihood function. 

Especially the outlier for \ch{CS2} at the resolution used by \citet{benneke_jwst_2024}, with a Bayes factor of $3\cdot10^4$, seems to be biased by the fact that the $4.6\micron$ absorption feature is by chance coinciding with two data points that have low uncertainties. Thus, these two points penalise models with low \ch{CS2} abundances more heavily than in an unbinned spectrum, leading to an overly tight posterior and, therefore, a jump in the Bayes factor. Only once the spectrum is sampled at a resolution high enough to not only probe the strength of the \ch{CS2} feature but also its width, the presence of the molecule can deduced confidently. However, such information cannot be obtained at R$\approx$50 in the case of \ch{CS2}.

Overall, our results stress that the most accurate representation of the spectrum for a retrieval is one without any binning, assuming that the
signal-to-noise ratio of the unbinned data points is still good enough to find absorption band of molecules.

\subsection{Testing for additional model complexity}
\label{sec:non_iso_tp}

In our fiducial model, we assumed an isothermal, cloud-free atmosphere. Isothermal atmospheres are often used in transmission spectroscopy, since the impact of a non-constant temperature profile can often not be constrained given the error bars of a typical JWST spectrum of a warm sub-Neptune. 

However, \citet{holmberg_possible_2024} used a variable temperature-pressure profile based on the analytic description by \citet{Madhusudhan_tp_2009}. In this section, we therefore test the impact of including this profile in our retrieval calculations. Additionally, we also test for the potential presence of clouds or hazes in the atmosphere of TOI-270 d.

\subsubsection{Non-isothermal atmosphere}

The previous study by \citet{holmberg_possible_2024} used a more complex temperature-pressure profile (T-p profile), following the study by \citet{Madhusudhan_tp_2009}. The use of a more complex T-p profile could be one reason for the differences between the retrieval results presented by \citet{holmberg_possible_2024} and ours.

\begin{figure}[ht]
  \centering
  \resizebox{0.8\hsize}{!}{\includegraphics{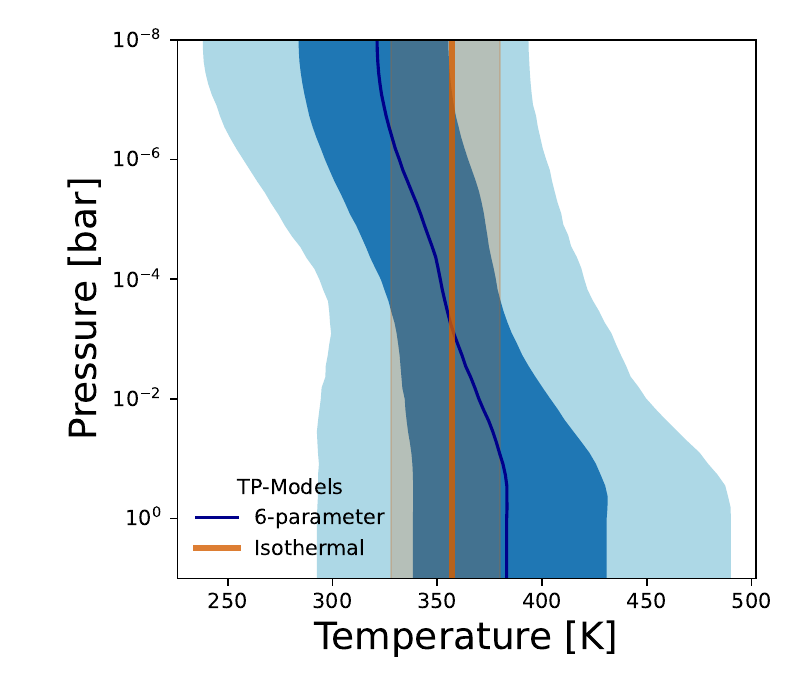}}
  \caption{Comparison of the two retrieved temperature-pressure profile models: the fiducial isothermal model and the more complex six-parameter model used in \citet{holmberg_possible_2024}. The blue-shaded areas refer to the $1\sigma$ and $2\sigma$ intervals, while the black, solid line represents the median temperature profile of the posterior sample. The red temperature profile corresponds to the median temperature and $1\sigma$ interval of our fiducial model.}
  \label{fig:tp-profile}
\end{figure}

The \citet{Madhusudhan_tp_2009} T-p profile is described by six free parameters and has been implemented as an alternative temperature profile description in \textsc{BeAR}. In our retrieval calculation employing this more complex description, we used the same prior distributions as listed in \citet{holmberg_possible_2024}. The resulting retrieved temperature profile is shown in Fig. \ref{fig:tp-profile}.

The T-p profile shows a slight increase in temperature from $320$ K at the top of the atmosphere to $380$ K at the bottom, at 10 bar. However, as the figure also suggests, this more complex profile is essentially consistent with the isothermal one from the fiducial model within $1\sigma$ throughout the whole pressure range.
Both retrievals lead to largely the same molecular abundance constraints, with only $\log x_{\text{\ch{H2O}}}=-3.8$ shifting by more than 0.1 from the fiducial value.

Since the five additional parameters do not contribute much to the overall outcome of the retrieval, it is disfavored with a Bayes factor of 0.25 compared to our fiducial model.

\subsubsection{Clouds and hazes}

In addition to our fiducial model, we also tested scenarios that include the impact of clouds and hazes. We tried two different cloud models implemented in \textsc{BeAR}: the two-parameter gray cloud model and the five-parameter non-gray cloud model based on the analytic fit to the general form of Mie efficiencies by \citet{kitzmann_nongreyclouds_2018}. The latter model can account for the transition of a gray cloud layer to a power-law haze as a function of particle size and wavelength. We did not use the cloud bottom as a free parameter but instead assumed that each cloud layer is extended by one atmospheric scale height below its top.

Both of these cloud model do not significantly affect the retrieval results. The resulting cloud-top pressures for the two models are $10^{-3.2 \pm 2.6}$ bar for the gray-cloud model and 
$10^{-1.5 \pm 2.5}$ for the \citep{kitzmann_nongreyclouds_2018} model. The median values for the (vertical) optical depths are $\tau = 10^{-1.8}$ for the former and about unity for the latter, respectively.

The contribution of the cloud decks are limited to wavelengths $<1.2\micron$ and are, therefore, only poorly constrained without the additional second order NIRISS spectrum. Thus, our spectrum can be well explained without the need for a thick cloud layer or hazes, making TOI-270 d one of the few sub-Neptunes that do not suffer strongly from a cloudy upper atmosphere, in line with predictions from \citet{brande_clouds_2024}. The studies by \citet{mikalevans_toi270_2023}, \citet{benneke_jwst_2024}, and \citet{holmberg_possible_2024} also found no strong evidence for the presence of clouds or hazes that influence the spectrum to a detectable degree.

\subsection{Testing for additional molecules}
\label{sec:retrieval_add_molecules}

In our fiducial model, we limited the molecules in the retrieval calculations to those used by \citet{benneke_jwst_2024}. In this section, we, therefore, test for the presence of additional molecules. In particular, we test for methyl chloride (\ch{CH3Cl}, also used by \citet{holmberg_possible_2024} and methyl fluoride (\ch{CH3F}). Our fiducial model also revealed potential constraints on sulfur species. In an additional model, we thus also include a broad selection of molecules that have been shown to play an important role in atmospheric sulfur chemistry \citep[e.g.][]{Moses2024DPS....5630806M}.

For the Bayesian model comparison a reduced version of the fiducial model without \ch{CO} and \ch{SO2} is used (see Appendix \ref{app:reduced_model}). This improves the Bayesian evidence to $-38404.41$, an improvement of the Bayes factor by a factor of 6.6 relative to the fiducial model. This way, we eliminate modifications to the Bayes factor that could originate from the previously undetected species.

\subsubsection{\ch{CH3Cl} and \ch{CH3F}}
\label{sec:ch3s_h2cs}

In this section we test two additional species of the methyl group, \ch{CH3Cl} and \ch{CH3F}. Both molecules have fairly similar absorption cross-sections in the wavelength range of the JWST observations that also overlap strongly with those of methane up to $\sim4\micron$. We include them one by one in separate retrieval calculation with \textsc{BeAR}, as well as all molecules that yielded abundance constraints from the fiducial model. The results for both retrievals are listed in Table \ref{table:post_defaultmodel}. The detailed posterior distributions and the posterior spectra are shown in Appendix \ref{sec:app_corner}.

As results presented in Table \ref{table:post_defaultmodel} and Figs. \ref{fig:corner_ch3cl} \& \ref{fig:corner_ch3f} clearly suggest, we obtain tight constraints for the abundances of both species. \ch{CH3Cl} is constrained with a mixing ratio of $10^{-3.17}$ while the methane abundance decreases to a value of $10^{-2.11}$, smaller by almost a factor of 2.5 compared to our reduced fiducial model. With a Bayes factor of > $10^4$, this model is decisively preferred over the latter.
Methyl fluoride yields almost the same posterior with a median mixing ratio of $10^{-3.2}$ but this time reducing the \ch{CH4} abundance only by a factor of about two to roughly $10^{-2}$. The Bayes factor for this model compared to our reduced fiducial one is about 2\,500 and, thus, lower than for the previous \ch{CH3Cl} case. However the Bayes factor of the \ch{CH3Cl} model over the \ch{CH3F} one is only about 6.4, indicating that the none of these two models is overwhelmingly preferred by the retrieval.

While in both models, the abundances of \ch{CS2}, \ch{NH3}, and \ch{CO2} do not differ much from those of the fiducial model, the water abundance increased strongly by more than one order of magnitude. With a mixing ratio of about $10^{-2.5}$ in both cases, the \ch{H2O} abundance is now compatible with an atmosphere that is enriched with 100 times compared to solar elemental abundances.

\subsubsection{Sulfur species}
\begin{figure*}[t!]
  \centering
  \resizebox{\hsize}{!}{\includegraphics{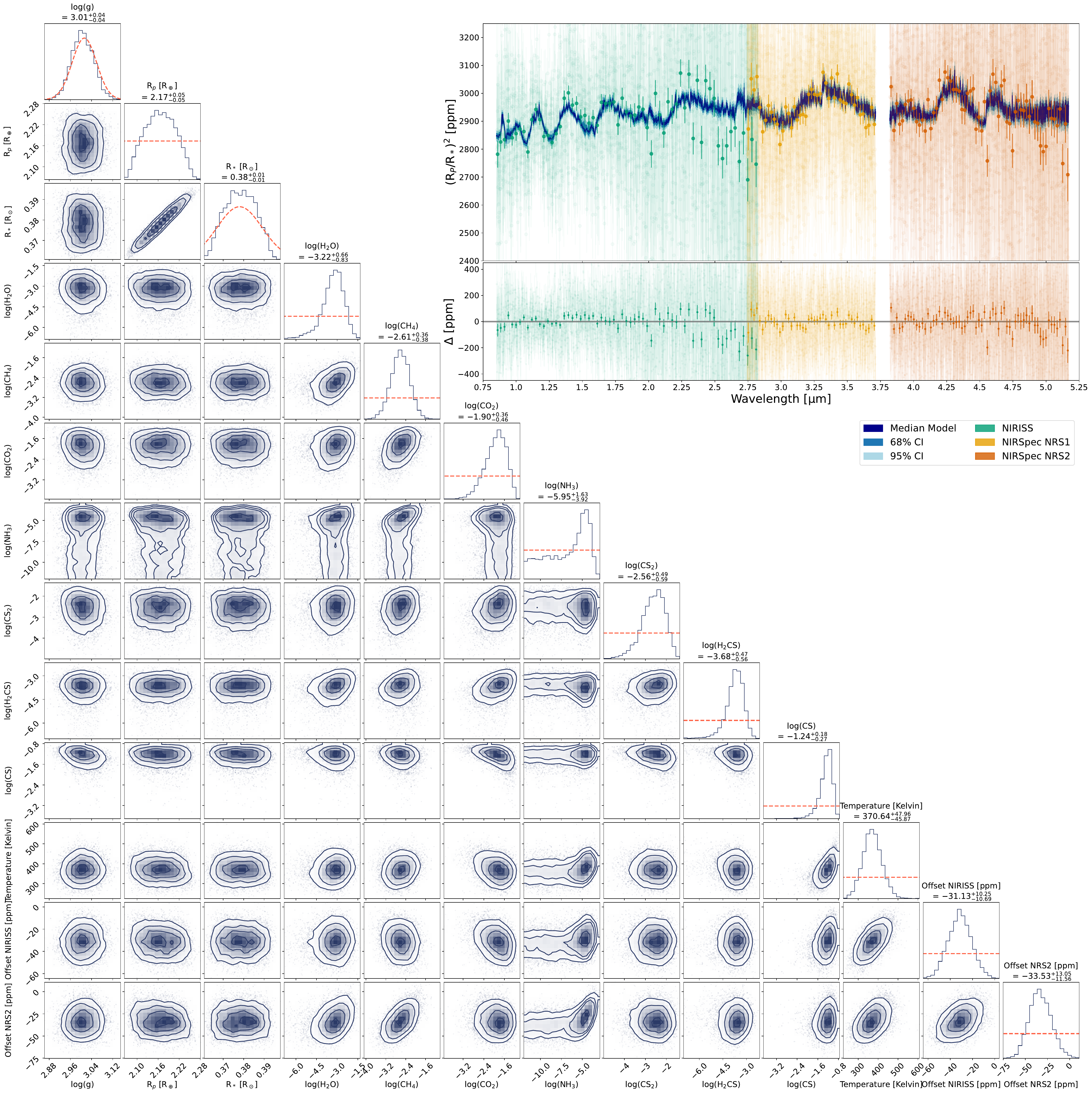}}
  \caption{Posterior distributions and transit spectra of our "sulfur" at the full spectral resolution of NIRISS SOSS GR700XD and NIRSpec G395H. The prior distributions for the retrieval parameters are shown in red. We also show a set of binned data points in the spectrum for readability, the 32-pixel binning.}
  \label{fig:corner_sulfur}%
\end{figure*}
Our fiducial model yields constraints on the abundance of carbon disulfide (\ch{CS2}) but did only provide upper limits for the presence of \ch{SO2}. Assuming a sulfur chemistry operating in the atmosphere of a warm sub-Neptune, the absence of the latter is to be expected. 

As described by \citet{Moses2024DPS....5630806M}, in the temperature regime of TOI-270 d of around 360 K, the usual very abundant hydrogen sulfide (\ch{H2S}) is easily destroyed, producing a wide range of sulfur species. Especially at higher metallicities, \ch{CS2} can be formed through a reaction network involving species, such as methyl (\ch{CH3}), thioformyl (HCS), monosulfide (CS), mercapto (SH), and thioformaldehyde (\ch{H2CS}). Other sulfur species that can be formed under these conditions are carbonyl sulfide (OCS) and sulfur mononitride (NS) \citep[][ Moses et al., in prep.]{Moses2024DPS....5630806M}.

We included all of these species where line lists are available in a retrieval calculation using \textsc{BeAR}. Of the species mentioned above, HCS does not seem to have published spectroscopic data and, therefore, could not be included as an opacity source in our retrievals. 

The species OCS, SH, \ch{H2S}, and \ch{CH3} only resulted in upper limits of about $10^{-5}$ in each case, while that of NS is $10^{-2}$. The absence of \ch{H2S} is easily explained by its destruction via chemical reactions as described above. Methyl and \ch{SH} are radicals and likely short-lived, such that they will not build up to detectable levels in the atmosphere. Carbonyl sulfide might also not be abundant enough to be detected. The posteriors for species with abundance constraints are listed in Table \ref{table:post_defaultmodel} and shown in Fig. \ref{fig:corner_sulfur}.

Our results suggest the presence of \ch{CS}, \ch{CS2}, and \ch{H2CS}. Especially carbon monosulfide is very abundant with a median volume mixing ratio of about 0.057, even higher than that of \ch{CO2}. Compared to the fiducial model, the abundance of methane decreased by about one order of magnitude, while the volume mixing ratio of \ch{H2O} increased by a factor of almost eight. Ammonia has a very non-Gaussian posterior distribution with a very long tail towards smaller mixing ratios. While its median value is about $10^{-5.9}$, the actual peak of the distribution is located at $x = 10^{-4.3}$ and, thus, closer to the values obtained in the previous retrieval calculations.

Overall, the detected species are consistent with the sulfur network discussed by \citet{Moses2024DPS....5630806M}. Whether our obtained mixing ratios are also realistic outcomes can only be tested by using a non-equilibrium chemistry model including an extensive sulfur reaction network. This, however, is beyond the scope of this work and will be addressed in future studies.

The very high CS abundance also results in a C/O ratio of 2.95, significantly larger than unity. The overall metallicity is increased by a factor of 100 compared to solar elemental abundances, while sulfur is enriched by a factor of more than 1000.

Another sulfur species that has been suggested by \citet{madhusudhan_carbon_2023} to be present in the atmosphere of the warm sub-Neptune K2-18 b is dimetyhl sulfide (\ch{(CH3)2S}, DMS). We test for the presence of this molecule by adding it as an additional species to the previous retrieval setup. The results are listed in Table \ref{table:post_defaultmodel}, while the posterior distributions and spectra can be found in Fig. \ref{fig:corner_sulfur_dms}. 

Based on our results, the spectrum seems to provide very minor evidence for the presence of \ch{(CH3)2S}. Similar to ammonia in the previous retrieval calculation, the posterior distribution for the mixing ratio of dimethyl sulfide shows a very long tail that moves the median value of the distribution of $10^{-6.2}$ away from its peak, located at roughly $10^{-4.8}$. The abundances of other species are essentially unaffected and show only minor deviations from the previous scenario without \ch{(CH3)2S}. This also reflected in the resulting Bayesian evidence that, with a value of -38398.92, is slightly smaller then the one of the retrieval calculation without this species (-38397.82), suggesting that there is no preference to include this species from a Bayesian perspective.

However, as discussed in Sect. \ref{chap:retrieval}, opacity data for \ch{(CH3)2S} is extremely limited. No line lists currently exist to calculate the line absorption. Only a small set of measured absorption cross-sections are available, taken at a pressure of one bar, most of which was \ch{N2}, and just three temperatures \citep{Sharpe2004ApSpe..58.1452S}. This scarcity makes it challenging to reliably assess a potential inference of \ch{(CH3)2S}.

\subsubsection{Degeneracies between \ch{CH3Cl}, \ch{CH3F}, and \ch{H2CS}}
\begin{figure}[h]
  \centering
  \resizebox{\hsize}{!}{\includegraphics{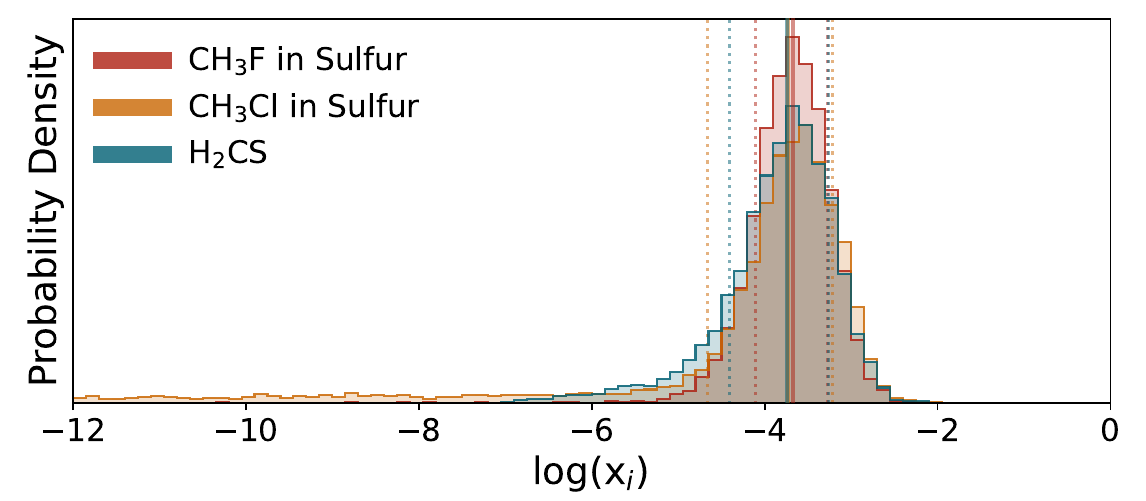}}
  \resizebox{\hsize}{!}{\includegraphics{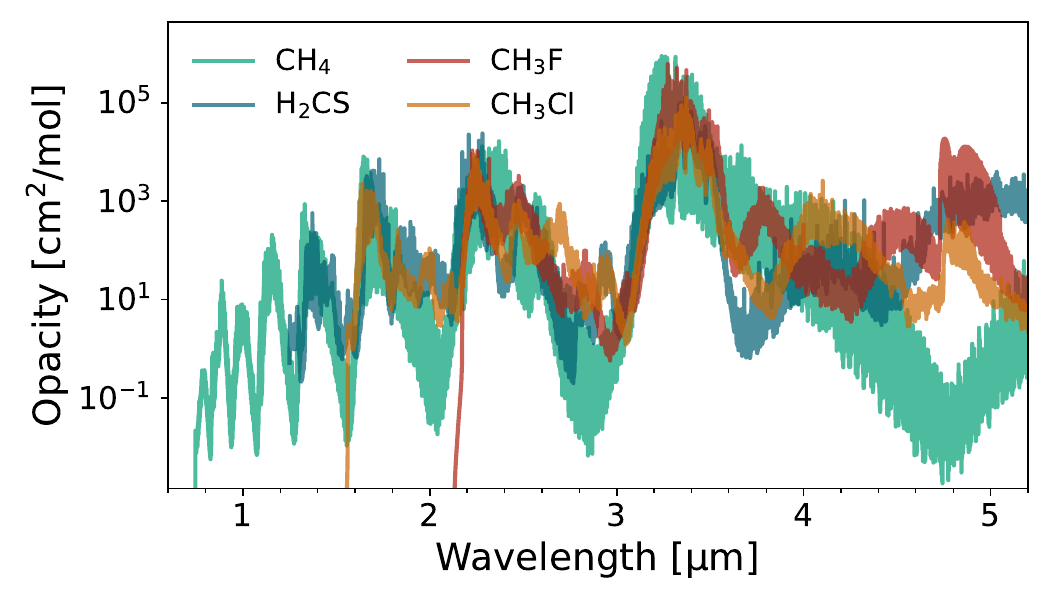}}
  \caption{Degeneracy of \ch{CH3Cl}, \ch{CH3F}, and \ch{H2CS} as potential model extensions.\textbf{Top}: Normalized posterior distributions for each molecule in the sulfur-species models. The posteriors are taken from the models listed in the last three columns of Table \ref{table:post_defaultmodel}. The solid lines represent the median value, while the dashed lines describe the corresponding $1\sigma$ intervals. \textbf{Bottom}: Comparison of absorption cross-sections as a function of wavelength at a temperature of 400 K and a pressure of 1 bar over the wavelength-range of the JWST observations used in this study.}
  \label{fig:other_molecules}
\end{figure}
Based on the results from the last two subsections, we obtain constraints on the abundance of \ch{CH3Cl}, \ch{CH3F}, or \ch{H2CS}. At the same time, the inclusion of these molecules also resulted in a decrease of the methane abundance in comparison to our fiducial model (see Table \ref{table:post_defaultmodel}). In this subsection, we test the inclusion of either \ch{CH3Cl} or \ch{CH3F} in our retrieval with the sulfur species from the preceding subsection to study if they impact the inference of any other species. The posterior results are listed again in Table \ref{table:post_defaultmodel}, with the full posterior distributions and spectra shown in Figs. \ref{fig:corner_sulfur_ch3cl} \& \ref{fig:corner_sulfur_ch3f} of Appendix \ref{sec:app_corner}.

By including these two species, we do not obtain a strong constraint on the abundance of the \ch{H2CS} anymore, but only an upper limit of about -4 for $\log x$. Most other species seem to be unaffected by the inclusion of either \ch{CH3Cl} or \ch{CH3F}, though. The Bayesian evidence listed in Table \ref{table:post_defaultmodel}, on the other hand, does not show a strong preference for either of these models, even though the one including methyl fluoride has the highest evidence value. In conclusion, the retrieval calculations are essentially unable to distinguish between these three different models based on the Bayesian evidence.
\begin{figure*}[t!]
  \centering
  \resizebox{\hsize}{!}{\includegraphics{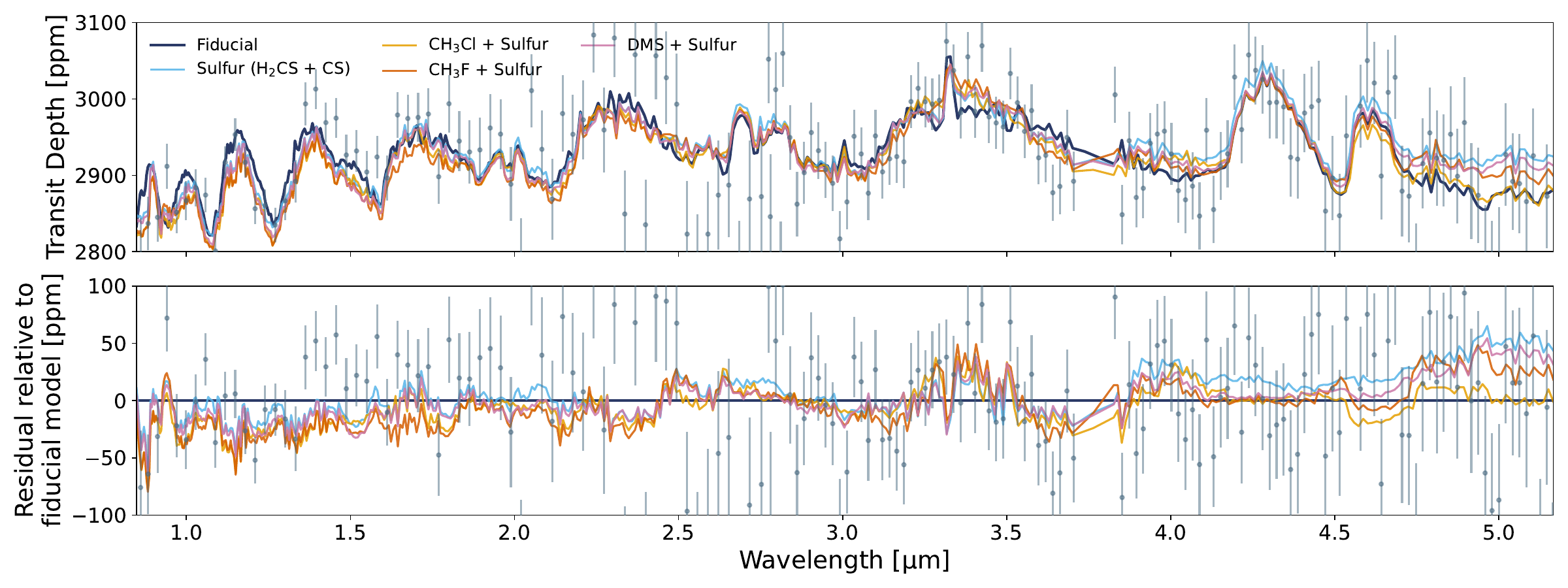}}
  \caption{Best-fit model spectra of the additionally tested molecules and their difference when compared to the fiducial atmosphere composition and the 32-pixel binned version of the transit spectrum. All three added species affect the best-fit spectrum similarly, improving the fit around the main methane feature at $3.4\micron$ and increasing transit depth at $4.0\micron$ and $4.8\micron$.}
  \label{fig:bestfit_ch3x_sulfur}
\end{figure*}

In the top panel of Fig. \ref{fig:other_molecules} we plot a comparison of the posterior distributions for all three species from based on their retrieval calculations including the sulfur species, while in the bottom panel we show the absorption cross-sections of these three species together with those of \ch{CH4}. The results clearly illustrate that the posteriors of \ch{CH3Cl}, \ch{CH3F}, and \ch{H2CS} are almost identical. This behavior can easily be explained by the corresponding absorption cross-sections shown in the bottom panel of Fig. \ref{fig:other_molecules}. All three species have the same basic band structure between 1 $\mu$m and about 4 $\mu$m, originating primarily from the C-H bonds of these molecules. Furthermore, they also largely overlap with the overall band structure of methane at these wavelengths. The latter behavior explains the decrease in the retrieved \ch{CH4} abundance when one of these three species is included in the model.

The only region where \ch{CH3Cl}, \ch{CH3F}, and \ch{H2CS} could be more easily distinguished from methane, is between 4.5 and 5.5 $\mu$m. However, all three of these molecules have strong absorption bands in this region, making it difficult to distinguish between them given the error bars of the JWST observation used in this study. Furthermore, this region also has significant contributions by \ch{CS2} and \ch{CO2} (see Fig. \ref{fig:bestfit_mols}), further complicating a clear distinction between \ch{CH3Cl}, \ch{CH3F}, and \ch{H2CS} at the given level of noise.

Figure \ref{fig:bestfit_ch3x_sulfur} shows the best-fit spectra of the four retrievals for \ch{CH3Cl}, \ch{CH3F}, and \ch{H2CS}, as well as our fiducial model. The spectra also again clearly demonstrate the similarities of the absorption bands of these three molecules and \ch{CH4} over the wavelength range of the JWST observations. Additionally, we also show the spectrum of the retrieval calculation including the sulfur species together with \ch{(CH3)2S}. The figure clearly illustrates that within the error bars of the observations, no clear distinction between the spectra is easily possible. 

Given the similarities between the opacities of \ch{CH3Cl} and \ch{CH3F}, other members of the methyl group could potentially have very similar cross-sections as well and, thus, roughly the same abundance constraints. This includes, for example, methyl bromide (\ch{CH3Br}) or methyl iodide (\ch{CH3I}). Unfortunately, line lists for these species are not available in the wavelength range of the JWST observations used here and additional observational data at longer wavelengths would be required to potentially detect them. 

Another possibility are hydrocarbons \citep[e.g.][]{niraula_titanasanexoplanet_2025} such as the photochemical product \ch{C2H6}, which was already included in the study of \citet{holmberg_possible_2024}. 
We find that in the near-infrared, the available opacity data is very limited compared to the other species used before. 
Both linelists \citep{hitran_2020} and measured absorption cross-sections \citep{C2H6_HITRAN} exist, but only cover wavelength ranges of $3.25-3.55\micron$ and $2.8-4.0\micron$ respectively. 
That is, the former is cut off at the lower edge of its main feature and the latter is limited in available temperatures and pressures. 
Another set of continuous ($1.61-5.55\micron$) ethane absorption cross-sections was measured by \citet{hewett_c2h6cross_2020}, but these are only usable outside the main feature around $3.4\micron$ due to saturation during the measurement.
Nonetheless, a more detailed investigation of photochemical products and their impact on the spectra of atmospheres containing significant abundances of methane is a potential avenue for future work.

\section{Conclusion and summary}
\label{section:discussion}

\subsection{The nature of TOI-270 d}

From our fiducial retrieval analysis, we find an atmosphere that is overall very similar to the previous result from \citet{benneke_jwst_2024} with respect to the resulting mean molecular weight and atmospheric temperature. However, we do also find significant differences to their work, especially in terms of the detailed chemical composition. In contrast, we are unable to reproduce the atmosphere presented by \citet{holmberg_possible_2024} with our own data reduction and retrieval framework. Their resulting atmosphere has a noticeably different T-p-profile and a much lower metallicity leading to a much smaller mean molecular weight.  

In all cases, including our non-fiducial models, we find water abundances at least an order of magnitude below the constraints from \citet{benneke_jwst_2024}. Our reduced water abundance and resultant lack of oxygen drives a comparatively high C/O ratio for all our retrieval models. Additionally, we are unable to confirm their initial signs of \ch{SO2}.

We find that our models with an expanded selection of molecules are decisively favored over the fiducial one in all cases. This suggests that the initial species selection by \citet{benneke_jwst_2024} is lacking some critical absorbers.
While the sulfur model could explain the high \ch{CS2} abundance \citep{Moses2024DPS....5630806M}, we cannot definitively rule out an atmosphere containing \ch{CH3Cl} or \ch{CH3F} due to the very similar opacities to \ch{H2CS} within our available wavelength range. It is surprising that the inclusion of either methyl compound essentially removes \ch{H2CS}. \citet{Leung_methyl_biosig_2025} recently proposed that \ch{CH3Cl} could potentially accumulate to detectable levels similar to our retrieved abundance. However, following their work this would require the presence of liquid water on TOI-270 d, which is unlikely considering our retrieved atmospheric temperature. Furthermore, our results should not be interpreted as evidence for a preference of the two methyl species. Their available opacity data is limited to wavelengths $>1.5\micron$ (\ch{CH3F}) and $>2\micron$ (\ch{CH3Cl}), see bottom panel of Fig. \ref{fig:other_molecules}. This does not cover the entire range of the NIRISS observation and, thus, could potentially give their retrieval calculations additional flexibility over \ch{H2CS} to fit the spectrum.

The species \ch{CS2}, \ch{H2CS}, and \ch{CS} are all part of a sulfur reaction network proposed by \citet{Moses2024DPS....5630806M} which would make this model a compelling scenario to explain the observational data. However, detailed non-equilibrium chemistry calculations with such a network need to be performed to ascertain if our abundance constraints are chemically plausible (Moses at al., in prep.). Especially the very high CS, that consequently results in a high C/O ratio, warrants a more in-depth analysis. It is possible, though, that this unusually high abundance is driven by instrument systematics or by certain choices made during the data reduction.

\begin{figure*}[ht]
  \centering
  \resizebox{\hsize}{!}{\includegraphics{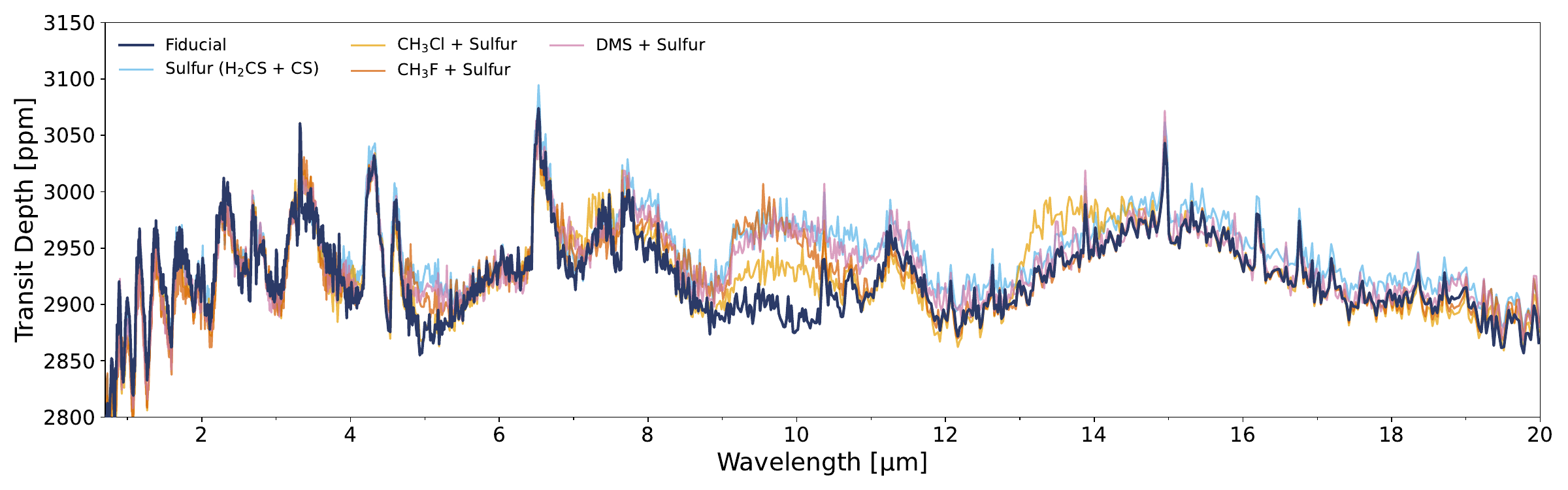}}
    \caption{Predicted transmission spectra of different retrieval scenarios. The best-fit models of each scenario have been used to generate spectra over a wider wavelength range, covering most of the JWST's full spectral range. }
  \label{fig:spectrum_jwst_range}%
\end{figure*}

Given the rather inconclusive results from the Bayesian framework that is unable to statistically distinguish the scenarios using either sulfur-species or methyl-group species, additional observations might be able to break this degeneracies. In Fig. \ref{fig:spectrum_jwst_range} we show spectra of the best-fit models for these scenarios over a wider wavelength range. The simulated spectra suggest that some these models could be distinguished with additional observations in the infrared region. The presence of \ch{CH3Cl}, \ch{CH3F}, and \ch{H2CS} would each lead to increased absorption around $10\micron$. While it appears challenging to disentangle these models, we already demonstrated in this work, that a retrieval is capable of significantly favoring the minor improvements from these models to the spectral fit over the fiducial model (see Fig. \ref{fig:bestfit_ch3x_sulfur}). Thus, it is plausible that the (combined) features at $8\micron$, $10\micron$, and $14\micron$ could be enough to resolve the present degeneracy through MIRI LRS or MRS observations.

The question of the structure of TOI-270 d below the probed atmosphere layer remains elusive.
While the potential signs of \ch{NH3} are more in line with a scenario without water condensation, 
the work by \citet{yang_chemical_2024} suggests that the increased \ch{CO2}/\ch{CH4} ratio across all retrievals would be indicative of a more water-enriched interior, while our lack of \ch{SO2} and especially \ch{OCS} would again imply the opposite according to their study. Additionally, for the sub-Neptunes LHS 1140 b and K2-18 b, \citet{huang_probing_2024} discussed, that an elevated C/O ratio could also be a consequence of oxygen depletion through condensation at higher pressures. The presence of a strong sulfur chemistry could be a crucial stepping stone into the characterization of the deeper atmosphere, as the retrieved molecular abundances could potentially be linked to the conditions in the deeper atmosphere through thermal quenching and the intensity of vertical mixing.

Clearly, there is a need for further photochemical and atmosphere-interior modeling in the search for the underlying explanation for this otherwise relatively well-characterized sub-Neptune.

\subsection{Data resolutions and Bayes factors}
\label{sec:bayes_factors}

Our investigation into using different data resolutions in a retrieval indicate that one cannot rely on native-resolution spectra without instrument specific changes to the forward model. Doing so could yield unpyhsical atmospheres in a retrieval calculation. While this can be avoided through data binning, this also leads to a loss of wavelength information and can introduce biases. Our approach of convolving the forward model with a Gaussian instrument profile based on the instrument resolution allows us to keep the maximal amount of information provided by the telescope data and retrieve identical or even more accurate compositions to retrievals using data that has been binned over multiple pixels.

We find that the spectral resolution used in atmospheric retrievals can be crucial in inferring the presence of species that do not dominate the shape of a spectrum over a large part of the spectral range, such as \ch{CH4}.
The Bayes factor in support of some molecule appears to be remarkably sensitive to the data resolution.
While it is consistently in favor of CS$_2$ and NH$_3$ in our case, for example, the difference between a noteworthy detection and weak evidence can change strongly. To avoid misinterpreting the presence or absence of molecules through the apparent volatility of the Bayes factor, JWST's precision should be fully utilized for suitable targets by analyzing their spectra at the highest possible resolution, even if this comes at higher computational costs and with the need to account for the LSF. 

\subsection{Summary}

Our main takeaways from this work are listed in the following:
\begin{itemize}
    \item We successfully reproduced the NIRISS and NIRSpec transit spectra of \citet{benneke_jwst_2024}, even though our data reduction and light curve fitting was performed at the native instrument resolution, demonstrating that there is no need to bin data for high S/N observations. 
    \item We utilize this more informative native resolution transit spectrum to achieve tighter constraints on the atmospheric composition of TOI-270 d, mostly aligning with the findings of \citet{benneke_jwst_2024}. We confidently detect \ch{CH4} and \ch{CO2} and find moderate evidence for \ch{CS2}.
    \item We identify either \ch{CH3F}, \ch{CH3Cl} or a rich sulfur chemistry including \ch{H2CS} and \ch{CS} as majorly favored extensions of the fiducial chemistry model, although the present data does not allow us to tell them apart.
    \item In our retrievals we come across unphysical atmosphere solutions when running retrievals without accounting for the LSF of the instrument. We show that this can be averted either by accounting for it in the forward model, or by binning the data at minimum every 16 pixels.
    \item The Bayes factor can show strong variability with data resolution for some molecules, indicating that care needs to be taken for the inference of molecules, whose signatures might be affected by the data resolution. The native resolution data should be most robust to such effects but leads to additional challenges in data reduction and light
    curve fitting.
    \item Atmospheric retrievals are capable of meaningfully probing for the presence of species with similar opacities as methane, such as CH$_3$X and \ch{H2CS}, even if they don't have exclusive absorption bands in our wavelength range, which makes telling them apart challenging.
\end{itemize}


\begin{acknowledgements}
We are very grateful to the referee for the careful reading of the paper and for his comments and detailed suggestions which helped us to improve the manuscript. 
This work is based on observations made with the NASA/ESA/CSA James Webb Space Telescope. The data were obtained from the Mikulski Archive for Space Telescopes at the Space Telescope Science Institute, which is operated by the Association of Universities for Research in Astronomy, Inc., under NASA contract NAS 5-03127 for JWST. These observations are associated with program GO 4098.

\end{acknowledgements}

\bibliographystyle{aa}
\setcitestyle{authoryear,open={(},close={)}}
\bibliography{references}

\begin{appendix}

\section{Eureka! settings summary}\label{section:eureka}

\begin{table}[h!]
\centering
\caption{Summary of the two Eureka! reductions, covering stages 1-3 of the pipeline.
\label{table:eureka}}
\begin{tabular}{lll}
\hline\hline
                            & NIRSpec     & NIRISS \\\hline
Data reduction stage 1      &             &        \\
Skipped steps               & \begin{tabular}[c]{@{}l@{}}IPC\\Persistence\end{tabular}& \begin{tabular}[c]{@{}l@{}}IPC\\Persistence\end{tabular}\\
Jump detection threshold    & $8\sigma$   & $8\sigma$\\
Bias-correction             & smooth      & smooth\\
Smoothing length            & 160         &  40   \\
Background subtraction      & yes ($2\sigma$) & no  \\\hline
Data reduction stage 2      &             &       \\
Performed steps             & \begin{tabular}[c]{@{}l@{}}srctype\\wavecorr\\extract\_2d\\extract\_1d\end{tabular}& \begin{tabular}[c]{@{}l@{}}srctype\\wavecorr\\extract\_2d\\extract\_1d\end{tabular}\\\hline
Data reduction stage 3      &             &       \\
Background subtraction      & yes ($3\sigma$) & yes ($3\sigma$)   \\
\begin{tabular}[c]{@{}l@{}}Spatial background\\outliers threshold\end{tabular}  & $3\sigma$ & $3\sigma$    \\
Spectral extraction half-width  & \begin{tabular}[c]{@{}l@{}}NRS1 4 pixels\\ NRS2 3 pixels\end{tabular} & 21 pixels \\
Background half-width       & 7 pixels & 28 pixels \\\hline

\end{tabular}
\end{table}

\section{NIRISS SOSS second-order data}\label{app:niriss_order2}
\begin{figure}[h]
  \centering
  \resizebox{\hsize}{!}{\includegraphics{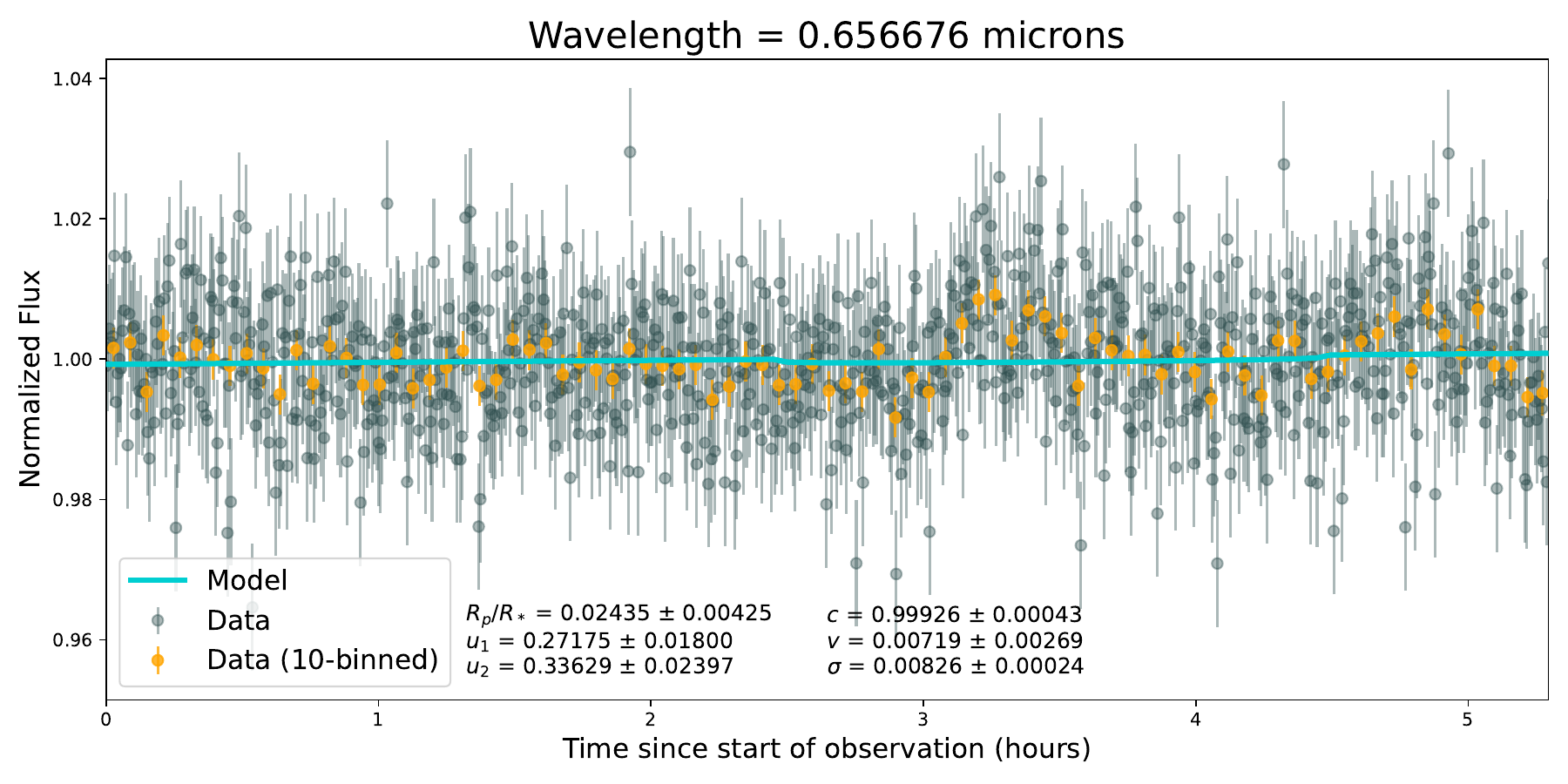}}
  \caption{Attempted light curve fit of the wavelength bin covering the H$\alpha$ spectral line at 656.46 nm. The fit is unsuccessful due to a increase in the stellar flux of $\approx1\%$ right in the center of the transit, leading to convergence of the transit depth to the lower prior boundary. Similar patterns starting at the 3 hour mark can be found all across the second order data, but none are as extreme as this one.}
  \label{fig:halpha}%
\end{figure}

When analyzing the second order NIRISS light curves at native resolution, we found that a large part of the spectrum showed signs of stellar activity.
Most prominently, the pixel column covering the H$\alpha$ spectral line shows an $\approx30$ minute-long event where the stellar flux increases by 1\%, completely hiding the planet's transit which is roughly 0.3\% deep.
The neighboring pixel columns show a similar increase at roughly half the amount.
When trying to fit these short-wavelength light curves we often converged towards the lower end of the planetary radius prior, suggesting that a more sophisticated model is required. Since this is a highly non-trivial effort, due to the fact that the shape is not consistent across different wavelengths, we instead chose to leave a detailed reduction of the second-order NIRISS data for future work.

One possibility is the procedure chosen by \citet{benneke_jwst_2024}, where the data is binned before the light curves are fitted, enhancing S/N and allowing for more efficient computation of a correction with Gaussian processes. The presence of this contamination might also be the cause of the larger-than-expected photometric scatter we obtain in the first-order data.

\section{NIRISS data beyond $2.5\micron$}
\label{app:niriss_beyond2.5}
\begin{figure}[h]
  \centering
  \resizebox{\hsize}{!}{\includegraphics{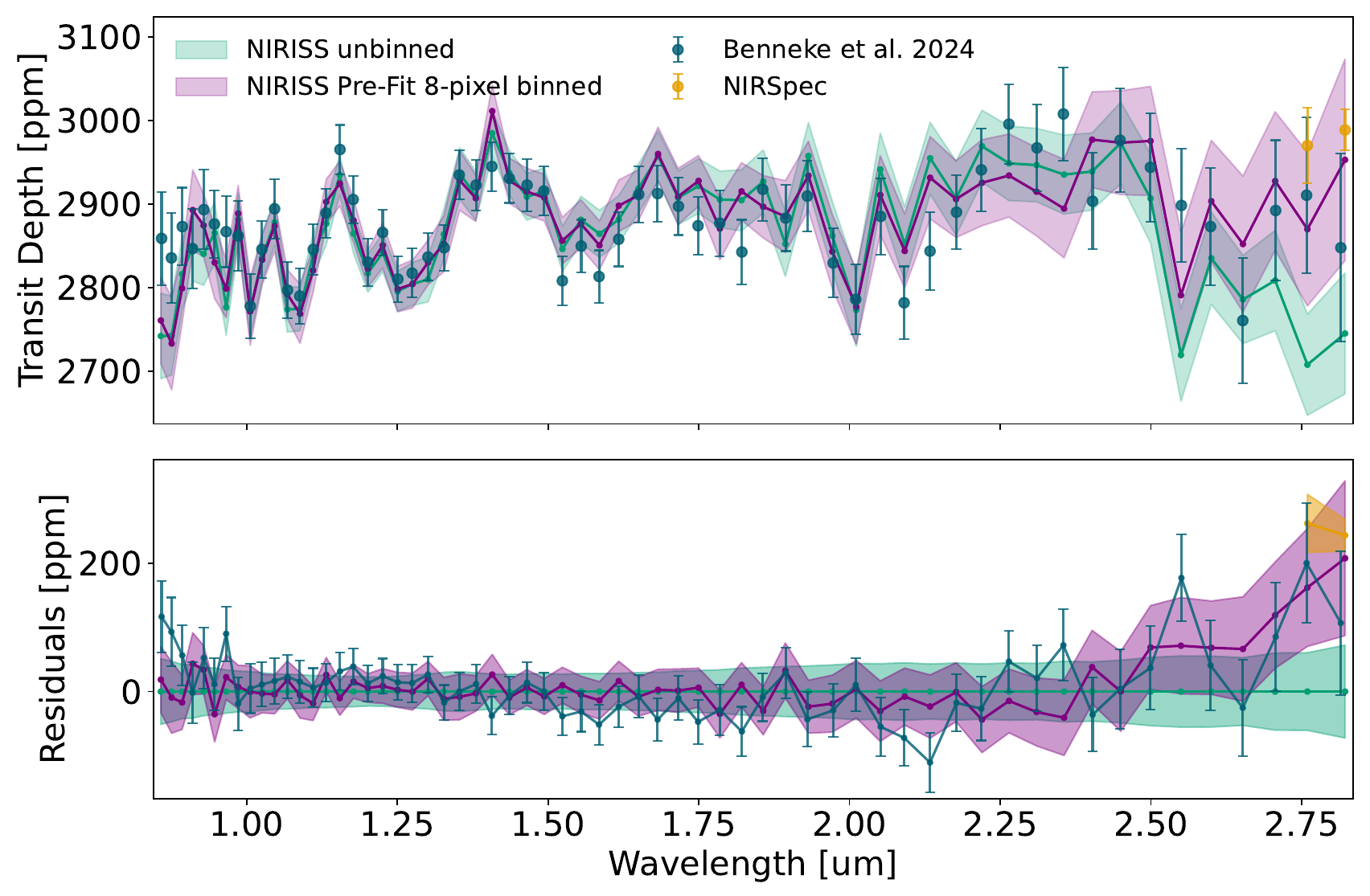}}
  \caption{NIRISS spectra based on two different choices for data binning. The top panel shows the resulting transit depths and compares them to the work by \citet{benneke_jwst_2024}. Bottom panel depicts the corresponding residuals.}
  \label{fig:niriss_8binned}%
\end{figure}

Our spectra from NIRISS SOSS Order 1 and NIRSpec NRS1 disagree at the overlapping region around $2.8\micron$. 
Additionally, the NIRISS spectrum's lower transit depth in this region also is in conflict with the measured depths of \citet{benneke_jwst_2024} and is generally not fit well in our atmospheric retrievals.
As shown in Fig. \ref{fig:niriss_8binned}, we find that this divergence originates in the spectral light curve fitting and can be avoided already if the data is binned every 8 pixels. We tested the sensitivity of our retrieval results to this reduced transit depth by running a retrieval with the >$2.5\micron$ NIRISS data cut off. Ultimately, only the retrieved \ch{H2O} and \ch{NH3} abundances are affected, the former increasing and the latter decreasing by an order of magnitude each.

\section{Fitting limb-darkening coefficients at high spectral resolution}\label{app:limb_darkening}

\begin{figure*}[h!]
  \centering
  \resizebox{\hsize}{!}{\includegraphics{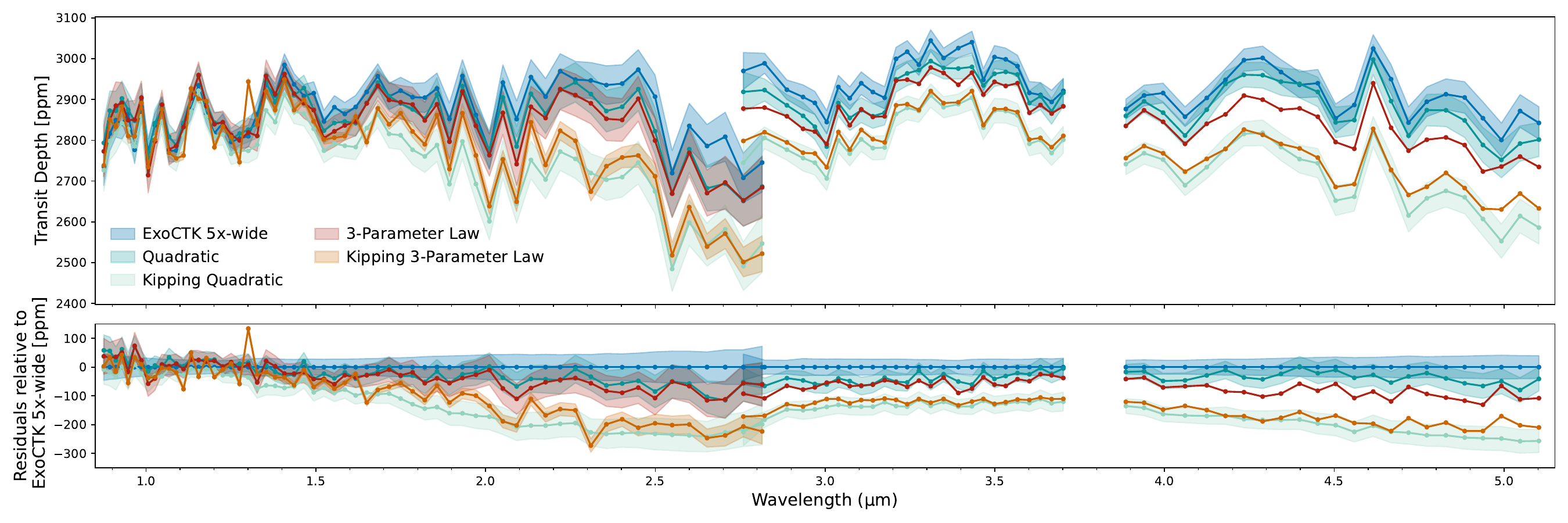}}
  \caption{Transit spectra for five different limb-darkening models. Spectra that have been fit using a re-parameterized limb-darkening law are biased towards smaller planet radii, depending on the S/N of the data. The spectra have been binned to R $\approx50$ for visual clarity.}
  \label{fig:data_limbdarkening}%
\end{figure*}

One of the challenges during the fitting of noisy light curves, which is typically the case when working at the native resolution of JWST data, is the estimation of limb-darkening parameters. We found that the suggested method by \citet{coulombe_ld_2024} of fitting the quadratic $u_1$ and $u_2$ parameters with wide uniform priors $\mathcal{U}\left(-3, 3\right)$ yielded rapidly varying parameters. These parameters sometimes resulted in unphysical light curves, often leading to limb-brightening.

On the other hand, using uniform priors $\mathcal{U}\left(0, 1\right)$ for $q_1$ and $q_2$ of the quadratic Kipping parameterization \citep{kipping_quadld_2013}, resulted in reasonable convergence but leads to an offset of up to 200 ppm that increases with data scatter compared to other quadratic limb-darkening choices.
This agrees with the results from \citet{coulombe_ld_2024}, who found a similar offset pattern that could be avoided by increasing the S/N through data binning.  In our broadband light curves, we found no difference between these two parameterizations, confirming that the Kipping-parameterization bias only appears in the low-S/N regime.

As a result, we chose to use a stellar model-informed approach, employing $5\times$-widened Gaussian priors based on the ATLAS stellar model \citep{castelli_atlas9_2004}.
While these models are possibly not entirely suitable for an M3V-dwarf, they nevertheless resulted in physically plausible light curves that were still allowed to depart from their priors.
Interestingly, the model-informed LDCs lead to spectra that are largely consistent with the free quadratic limb-darkening case, but without the need for unphysical limb-darkening solutions. The model-informed priors also produce the closest match with the spectra of the two previous studies by \citet{benneke_jwst_2024} and \citet{holmberg_possible_2024}.

Following the work by \citet{keers_ld_2024}, we also calculated transit spectra using the three-parameter law by \citet{sing_3param_2009} in its normal form as well its modified Kipping variant \citep{Kipping_3param_2015}.
Here, we obtained similar results to the quadratic LDCs: rapidly varying and partially unphysical coefficients for the default Sing law and a bias towards smaller planet radii for the re-parameterized version.
The resulting spectra using the Sing law are largely in agreement with those for the quadratic law, yet they do yield much smaller uncertainties in the transit depth, especially for the NIRSpec data.

\section{Light curve fit diagnostics}
\label{app:lc_diagnostics}

\begin{figure}[h]
  \centering
  \resizebox{0.49\hsize}{!}{\includegraphics[trim={1cm 0 0 0}]{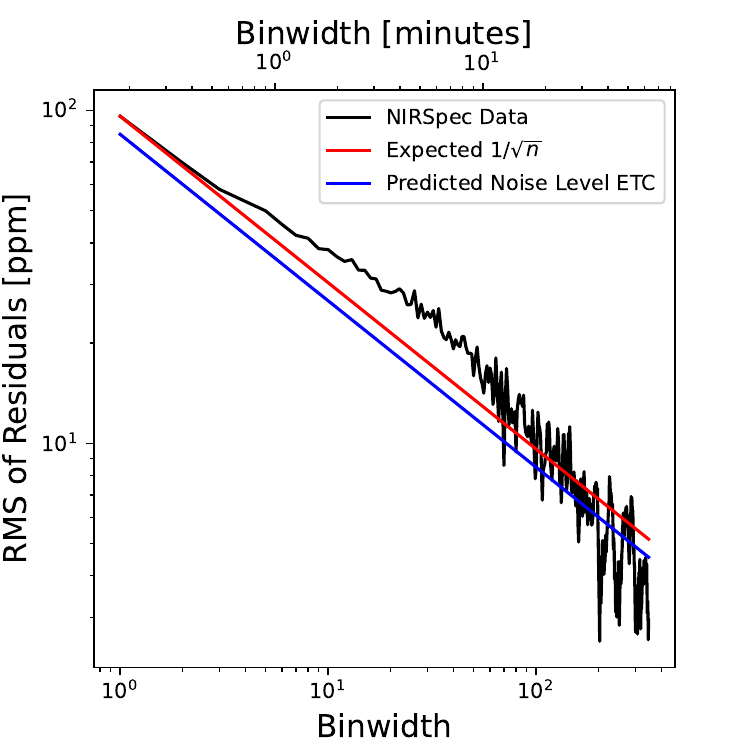}}
  \resizebox{0.49\hsize}{!}{\includegraphics[trim={1cm 0 0 0}]{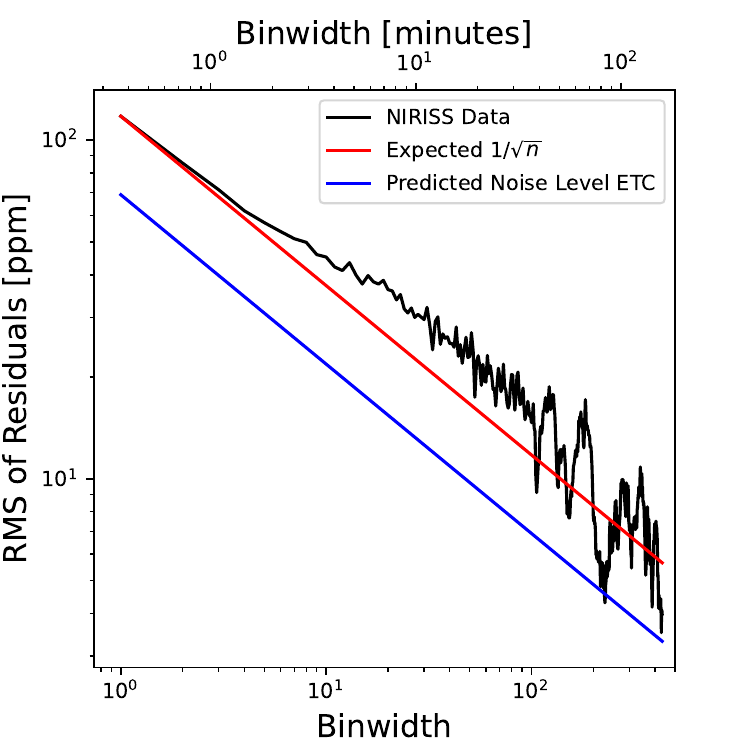}}
  \caption{Allan-variance of the broadband data. \textbf{Left:} NIRSpec data, NRS1, and NRS2 combined. \textbf{Right:} NIRISS data.}
  \label{fig:allan_broad_nirspec}%
\end{figure}

In Fig. \ref{fig:allan_broad_nirspec} we show the broadband Allan-variance plots, corresponding to the broadband light curves from Figure \ref{fig:broadband_lcs} \citep{allan}. For the NIRSpec data in the left panel, we identified some correlated noise on timescales shorter than the transit event. In case of the NIRISS observation (right panel) we found correlated noise on all timescales compared to the expected $1/\sqrt n$-scaling.

\FloatBarrier

\section{Evaluating planet masses with abundance constraints}
\label{chap:mass_comparison}
\begin{figure*}
        \centering
        \resizebox{\hsize}{!}{\includegraphics{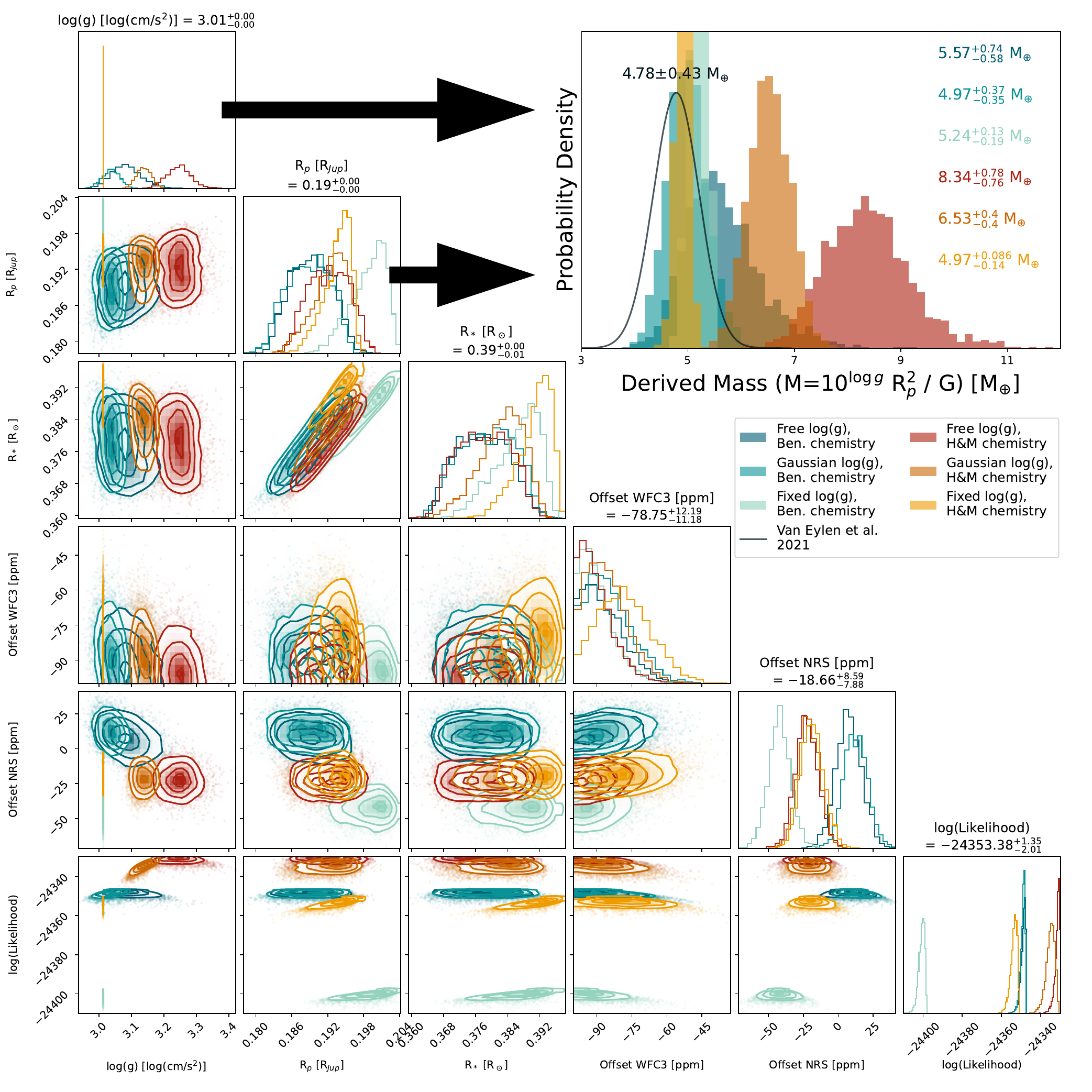}}
        \caption{Posterior distributions and derived masses of the six retrieval cases described in section \ref{chap:mass_comparison}. The retrievals were performed with fixed chemical abundances, temperatures and mean molecular weights as reported in \citet{benneke_jwst_2024} and \citet{holmberg_possible_2024}. We can enforce a planet mass consistent with \citet{vaneylen_toi270_2021} for either atmospheric composition through the use of a delta-prior, but this also forces a change in the radii of both the planet and the star. All retrievals favor a planet mass on the larger end of the RV-measurement predictions.}
        \label{fig:mass_comparison}%
    \end{figure*}

As a first test of the two previously published results, we analyze their retrieved atmospheric composition for consistency with our retrieval framework \textsc{BeAR}. We ran three retrievals for both sets of chemical abundances, mean molecular weights, and atmospheric temperatures, which are fixed to the reported values. 
In the case of \citet{holmberg_possible_2024}, the mean molecular weight was not reported and is instead inferred from the reported abundances of the One-Offset case as 3.1 amu.

This results in five remaining free parameters: the surface gravity of the planet ($\log(g)$), the planet radius ($R_\mathrm{p}$), the stellar radius ($R_*$), and the two offset parameters between the three datasets.
The priors of these fit parameters are the same as we use in our atmospheric retrievals, except that the first offset is set between the NRS1 and WFC3 data instead.

By combining the retrieved planet radius and surface gravity, we can use the measured atmospheric scale height to constrain the planet mass depending on the atmospheric composition. The retrievals were run on the published NIRSpec transit spectrum from \citep{holmberg_possible_2024}, combined with the Hubble WFC3 data from \citet{mikalevans_toi270_2023}. Naturally, this should bias this analysis towards the abundances from \citet{holmberg_possible_2024}, as they were retrieved from this exact same data while the abundances from \citet{benneke_jwst_2024} were derived from a different reduction of the NIRSpec data combined with the more recent NIRISS observation. 

We find, that the higher mean molecular weight in \citet{benneke_jwst_2024} seems to largely agree with the radial velocity mass constraints from \citet{vaneylen_toi270_2021}, although they all end up on the higher end of the mass range.
On the other hand, the lower-metallicity composition from \citet{holmberg_possible_2024} can only be brought into agreement with the independent mass measurement through a delta prior on the surface gravity. However, even in that case the planet and stellar radii start to compensate for the otherwise disfavored scale height.
The same effect is visible when we fix the surface gravity in the other compositional scenario.
Overall, the Bayesian evidence still heavily favors the two high-mass models over the mass-consistent models.
This indicates that if the mass, or equivalently the surface gravity, of the planet is not considered before running a retrieval, one might arrive at a more flexible but physically implausible solution.
With inclusion of a heavier, weakly absorbing molecule, such as N$_2$, one could increase the mean molecular weight of the atmosphere to the same value as seen in \citet{benneke_jwst_2024} without affecting the spectrum too much besides the scale height, but no inclusion of such a background gas was mentioned in \citet{holmberg_possible_2024}.

\section{Line list data sources for opacity calculations}
\label{sec:app_opacities}

\begin{table}[h!]
\caption{Line list sources and databases for the opacity calculations of the used chemical species.
\label{table:linelists}}
\centering 
\begin{tabular}{lll}
\hline \hline
Molecule   & Reference & Database \\
\hline
\ch{H2}-\ch{H2} CIA  &\citet{Abel_doi:10.1021/jp109441f} & HITRAN\\
\ch{H2}-\ch{He} CIA  &\citet{Abel2012JChPh.136d4319A} & HITRAN\\
\ch{CH4}   &\citet{ch4} & ExoMol\\
\ch{H2O}   &\citet{h2o} & ExoMol\\
\ch{CO2}   &\citet{co2} & ExoMol\\
\ch{CO}    &\citet{co} & ExoMol\\
\ch{NH3}   &\citet{nh3_1} & ExoMol\\
\ch{CS2}   &\citet{hitran_2020} & HITRAN\\
\ch{SO2}   &\citet{so2} & ExoMol\\
\ch{H2S}   &\citet{h2s} & ExoMol\\
\ch{OCS}   &\citet{hitran_2020} & HITRAN\\
\ch{(CH3)2S} &\citet{Sharpe2004ApSpe..58.1452S} & HITRAN\\
\ch{CH3Cl}   &\citet{Owens2018MNRAS.479.3002O} & ExoMol\\
\ch{CH3F}    &\citet{Owens2019PCCP...21.3496O} & ExoMol\\
\ch{H2CS}    &\citet{h2cs} & ExoMol\\
\ch{CS}      &\citet{cs} & ExoMol\\
\ch{NS}      & \citet{hs_ns} & ExoMol\\
\ch{SH}      &\citet{hs_ns} & ExoMol\\
\ch{CH3}     &\citet{ch3} & ExoMol\\
\hline
\end{tabular}
\end{table}

\section{Reduced fiducial model results}
\label{app:reduced_model}
The reduced version of the fiducial model does not include \ch{CO} and \ch{SO2} as opacity sources, as they remained unconstrained in the initial retrieval. Table \ref{table:post_reducedmodel} contains the parameter constraints retrieved with this model, similar to Table \ref{table:post_defaultmodel}.

\begin{table}[t]
\begin{center}
\renewcommand*{\arraystretch}{1.3}
\caption{
Median values and $1\sigma$-intervals from the posterior distributions for the retrieval calculations of TOI-270 d using the reduced version of the fiducial model. 
\label{table:post_reducedmodel}
}
\begin{tabular}{lc}
\hline \hline
Parameter                & Retrieved value\\ 
\hline
$\ln \mathcal Z$         & -38404.41                \\
\textbf{Retrieved}       &                          \\
$R_\mathrm{p}$ (\Rearth) & $2.16 \pm 0.05$          \\
$T_\mathrm{p}$ (K)       & $399^{+35}_{-36}$        \\
$\log x_{\ch{H2O}}$      & $-3.41^{+1.26}_{-5.29}$  \\
$\log x_{\ch{CH4}}$      & $-1.58^{+0.24}_{-0.26}$  \\
$\log x_{\ch{CO2}}$      & $-1.27^{+0.16}_{-0.18}$  \\
$\log x_{\ch{NH3}}$      & $-3.21^{+0.38}_{-0.55}$  \\
$\log x_{\ch{CS2}}$      & $-1.98^{+0.34}_{-0.52}$  \\
$\Delta_\mathrm{NIRISS}$ (ppm)  & $-36.3 \pm 9.7$   \\
$\Delta_\mathrm{NRS2}$ (ppm)   & $-7.6 \pm 11.8$    \\
\hline
\end{tabular}
\end{center}
\end{table}

\FloatBarrier

\section{Posterior distributions and spectra}
\label{sec:app_corner}

In the following, we show all posterior distributions and spectra from the models of Table \ref{table:post_defaultmodel} that are not present in the main text. All figures are set up identical to Fig. \ref{fig:corner_defaultmodel}.

\begin{figure*}[b!]
  \centering
  \resizebox{\hsize}{!}{\includegraphics{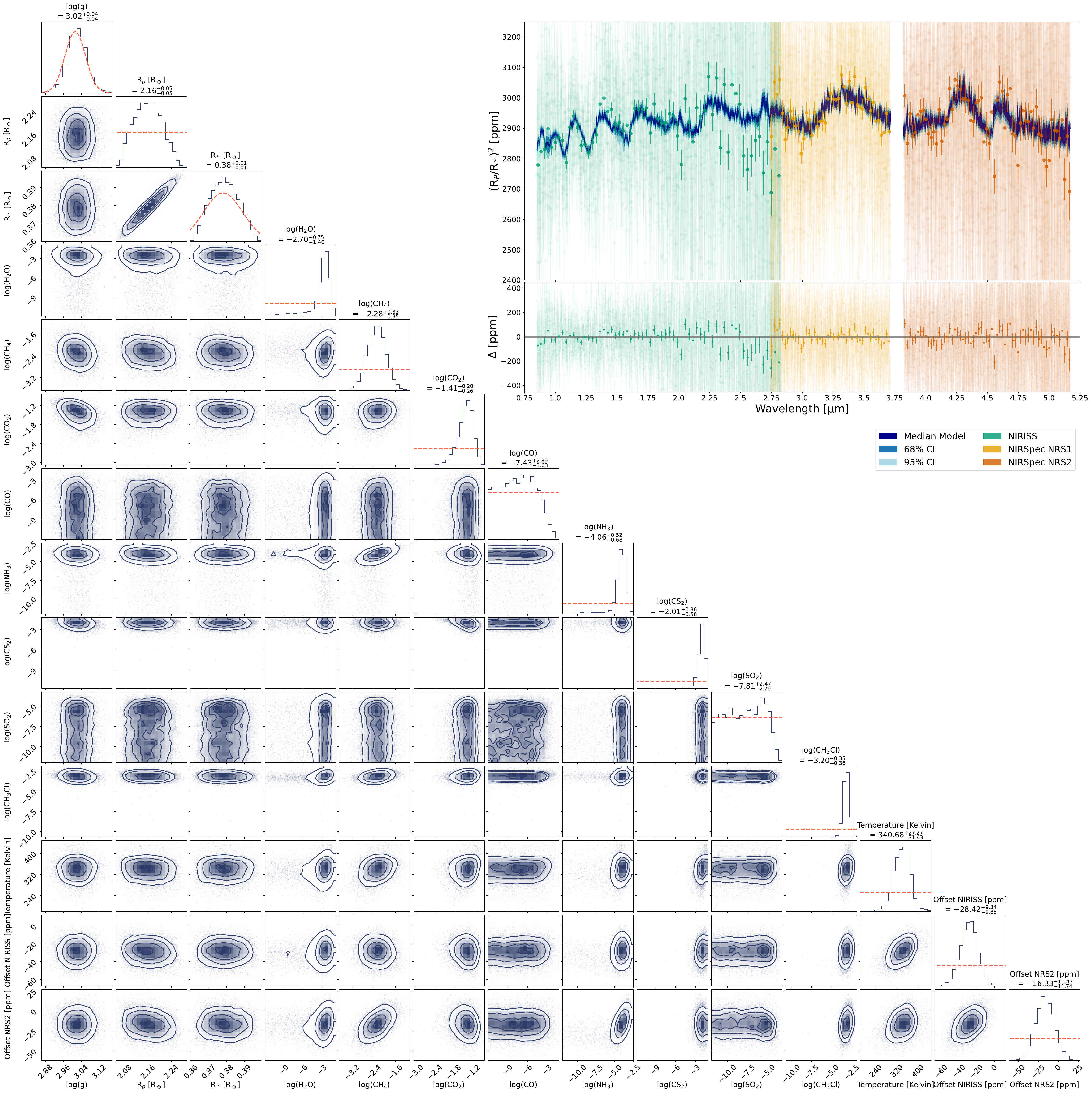}}
  \caption{Parameter posteriors and model spectra for the forward model including \ch{CH3Cl}.}
  \label{fig:corner_ch3cl}%
\end{figure*}

\begin{figure*}[ht]
        \centering
        \resizebox{\hsize}{!}{\includegraphics{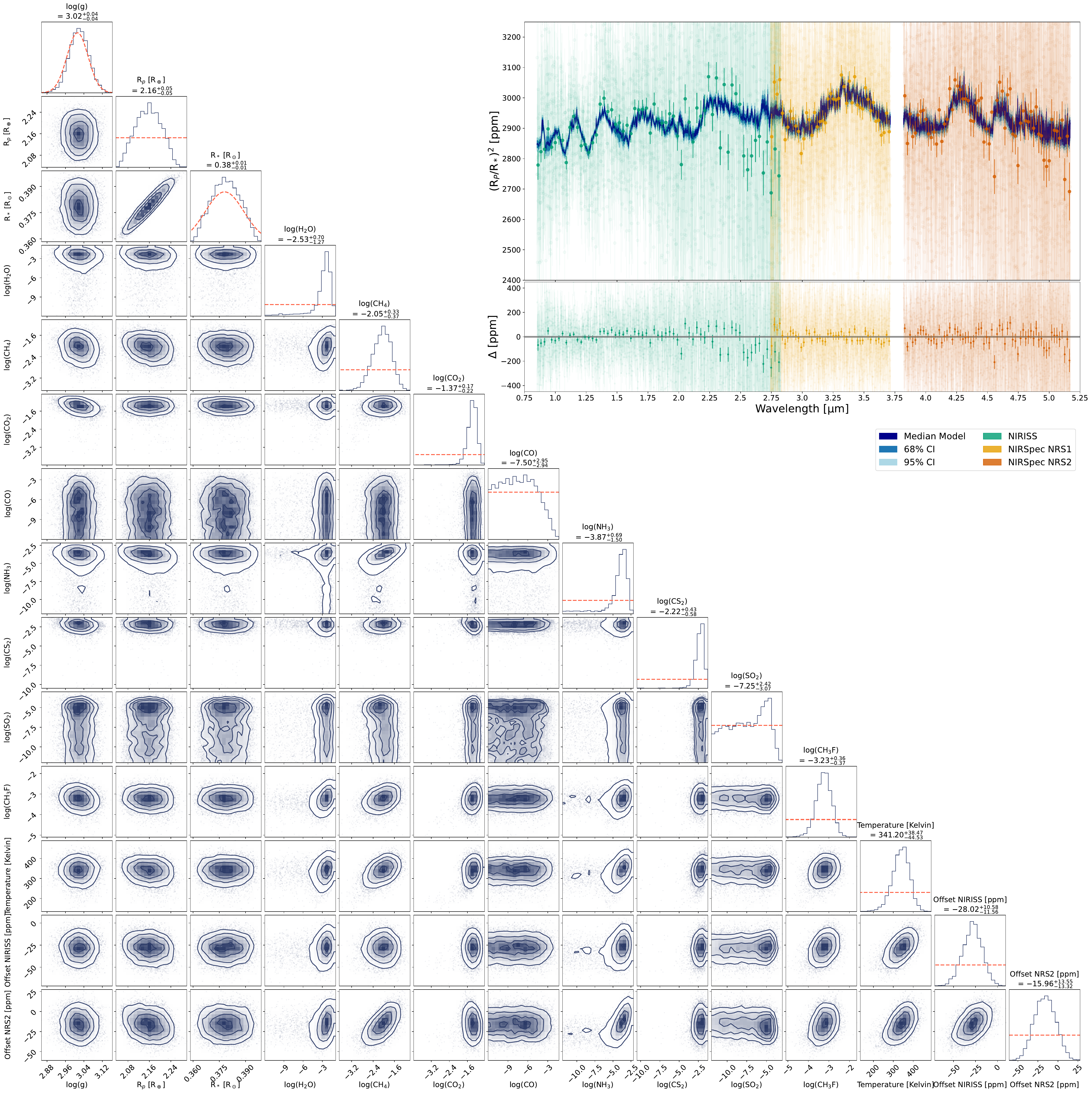}}
        \caption{Parameter posteriors and model spectra for the forward model including \ch{CH3F}.}
        \label{fig:corner_ch3f}%
    \end{figure*}

\begin{figure*}
        \centering
        \resizebox{\hsize}{!}{\includegraphics{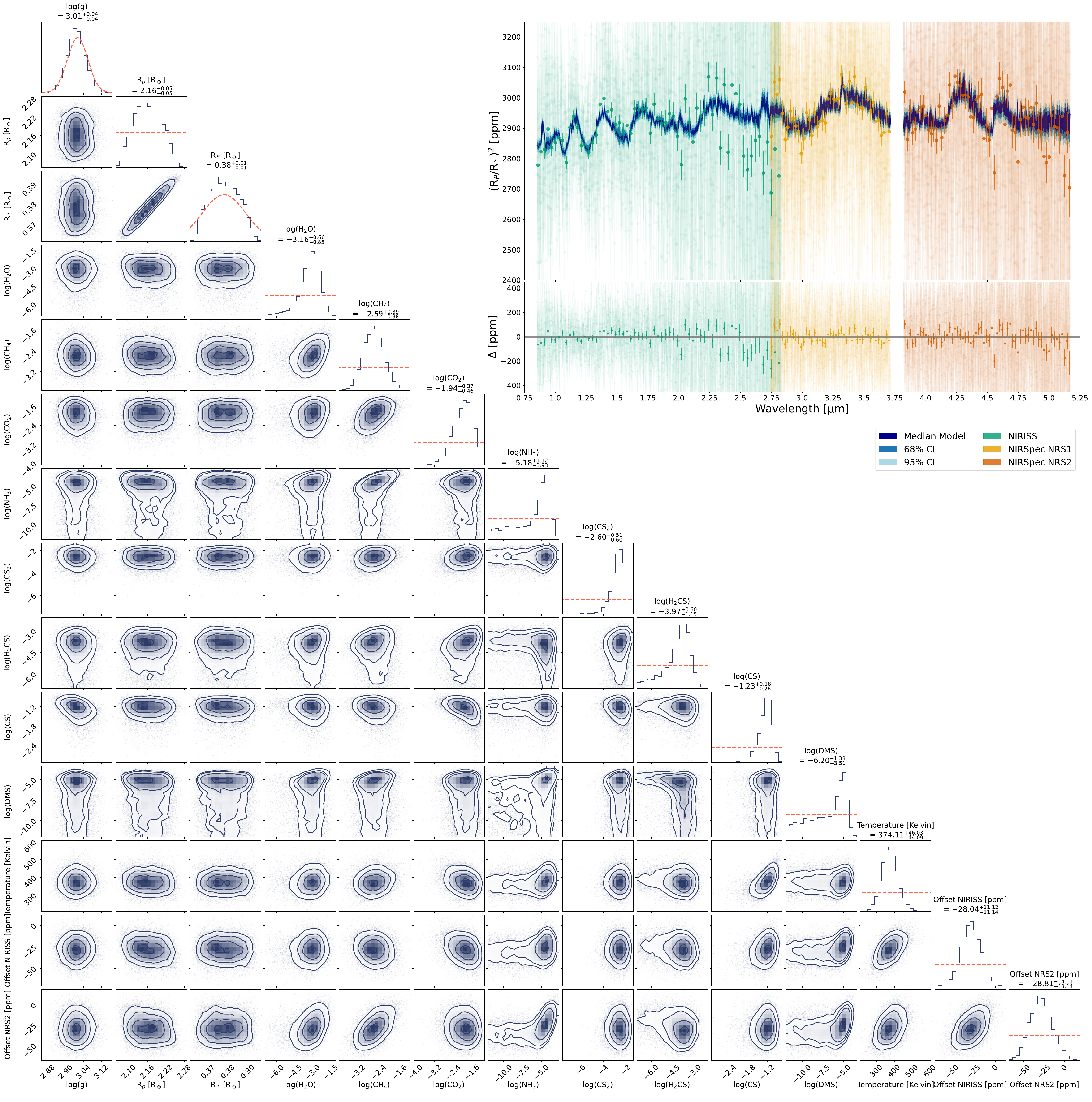}}
        \caption{Parameter posteriors and model spectra for the sulfur forward model including the sulfur species and \ch{(CH3)2S}}.
        \label{fig:corner_sulfur_dms}%
    \end{figure*}

\begin{figure*}[ht]
        \centering
        \resizebox{\hsize}{!}{\includegraphics{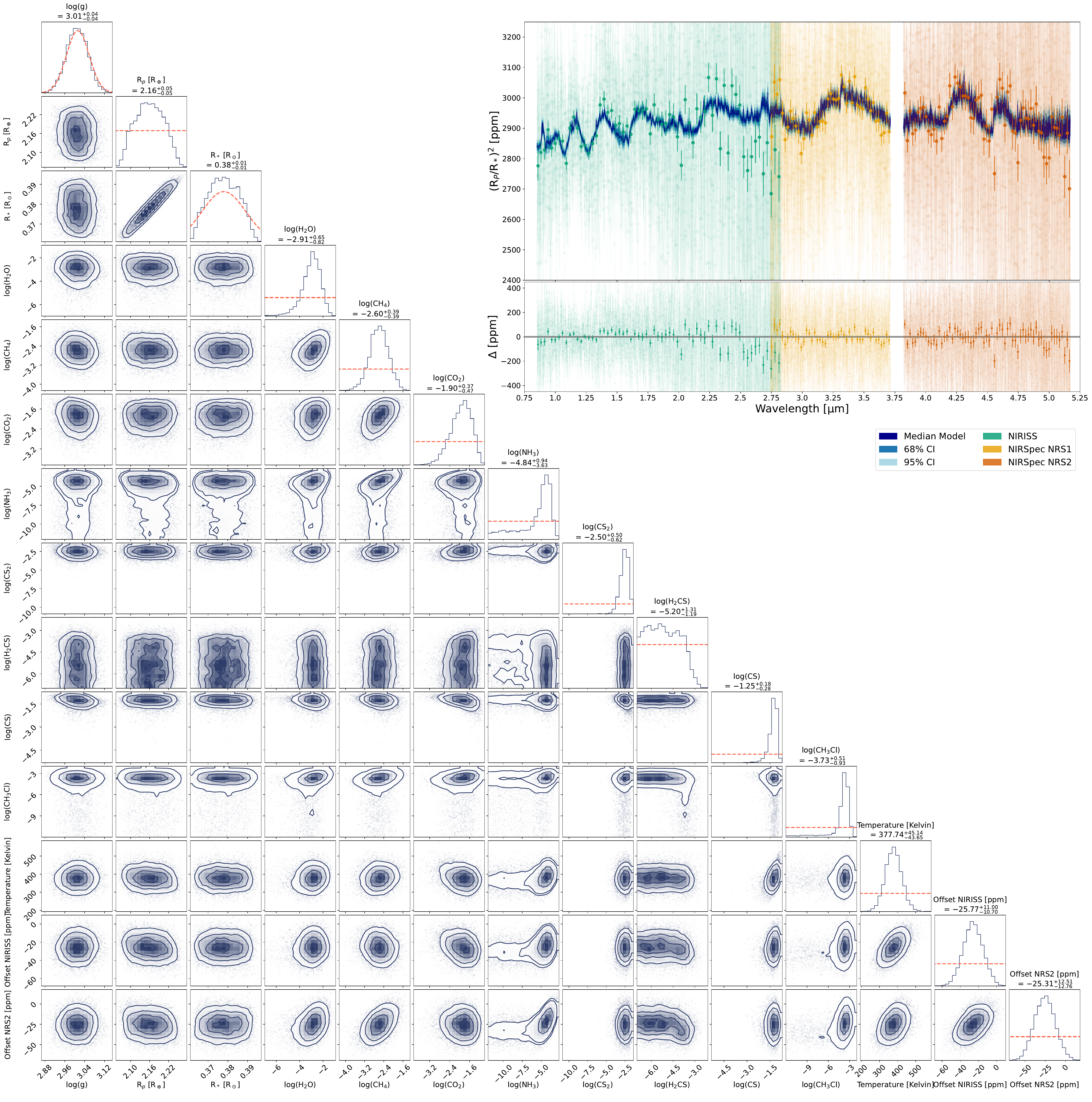}}
        \caption{Parameter posteriors and model spectra for the sulfur forward model including \ch{H2CS} and CS and \ch{CH3Cl}.}
        \label{fig:corner_sulfur_ch3cl}%
    \end{figure*}

\begin{figure*}[ht]
        \centering
        \resizebox{\hsize}{!}{\includegraphics{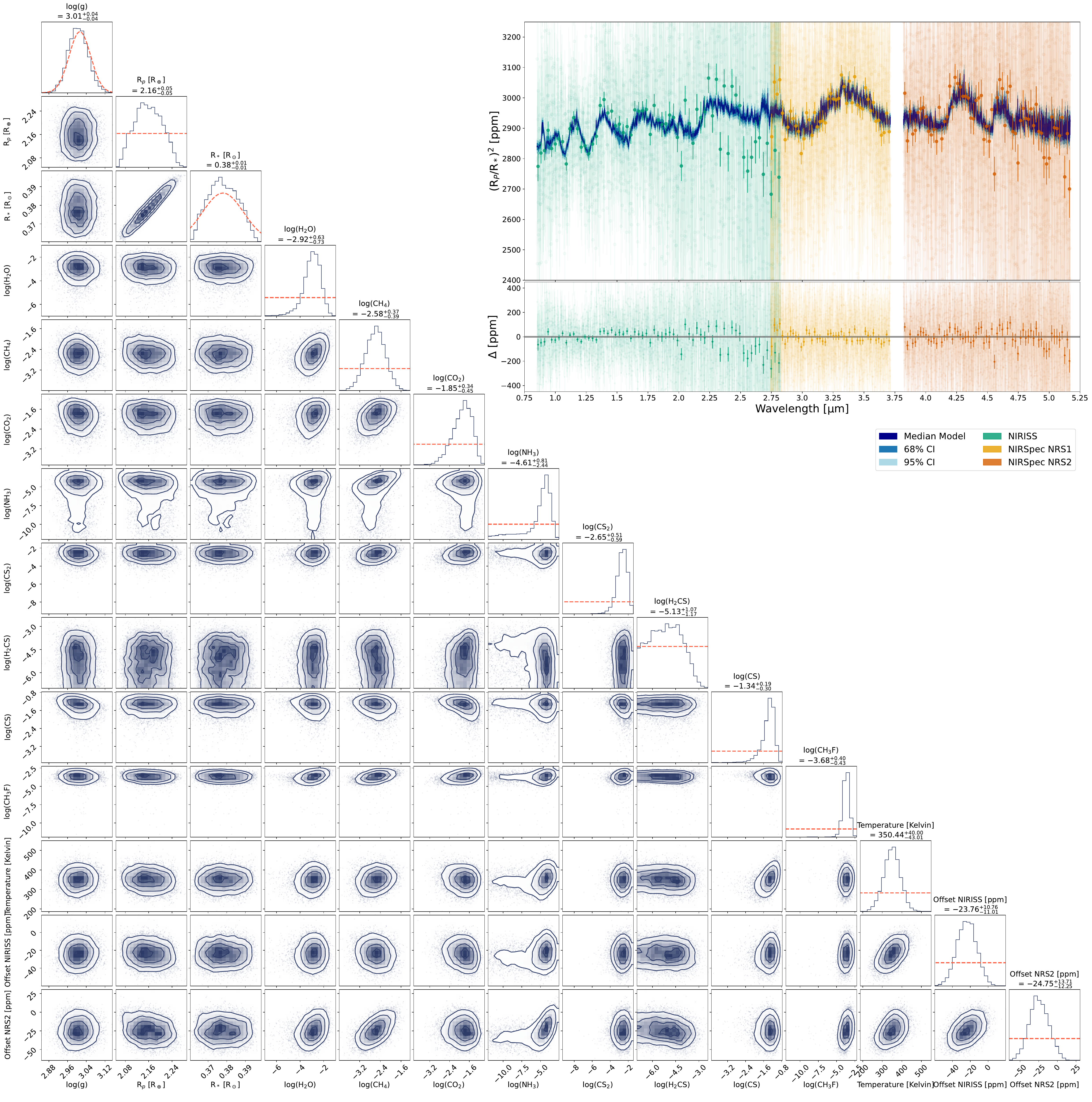}}
        \caption{Parameter posteriors and model spectra for the sulfur forward model including \ch{H2CS} and CS and \ch{CH3F}.}
        \label{fig:corner_sulfur_ch3f}%
    \end{figure*}

\clearpage

\end{appendix}

\end{document}